\documentclass[aps,preprint]{revtex4}%
\usepackage{amsfonts}
\usepackage{amsmath}
\usepackage{amssymb}
\usepackage{graphicx}%
\providecommand{\U}[1]{\protect\rule{.1in}{.1in}}

\ifx\pdfoutput\relax\let\pdfoutput=\undefined\fi
\newcount\msipdfoutput
\ifx\pdfoutput\undefined\else
\ifcase\pdfoutput\else
\ifx\paperwidth\undefined\else
\ifdim\paperheight=0pt\relax\else\pdfpageheight\paperheight\fi
\ifdim\paperwidth=0pt\relax\else\pdfpagewidth\paperwidth\fi
\fi\fi\fi
\begin{document}
\preprint{ }
\title[Duality violation]{Duality violation from a grating}
\author{Daniel Mirell}
\affiliation{Department of Chemistry, University of California, Irvine, CA 92697,
Electronic address: dmirell@uci.edu}
\author{Stuart Mirell}
\affiliation{Department of Radiological Sciences, University of California, Los Angeles, CA
90024, Electronic address: smirell@ucla.edu }
\author{}
\affiliation{}

\begin{abstract}
Diffraction orders in the continuous wave regime generated by a Ronchi
transmission grating in a standard threshold configuration are shown
theoretically to violate quantum duality for a locally real representation.
The phenomenon superficially resembles Rayleigh anomalies but is notably
distinguished from those anomalies by a prediction of probability
non-conservation. This prediction is experimentally tested with a 633 nm laser
beam at normal incidence on gratings giving that threshold condition for the
$\pm3^{rd}$ order pair. Transient intersection of the $0^{th}$ order with an
independent 633 nm laser beam demonstrates a duality-violating probability
non-conservation in good agreement with the theoretical prediction.

\end{abstract}
\volumeyear{year}
\volumenumber{number}
\issuenumber{number}
\eid{identifier}
\date{July 6, 2011}
\startpage{1}
\endpage{ }
\maketitle

\section{$^{{}}$Introduction}

The compact mathematical formalism of quantum mechanics developed in the late
1920's continues as an extraordinarily successful representation of physical
phenomena. The very compactness itself is widely perceived as evidence of the
validity, beauty, and completeness of that formalism. This perception has
persisted despite the evolving understanding over the years that strict
adherence to the formalism imposes fundamental violations of classical
physical reality. The resolution of these violations, given the constraints of
that strict adherence, was understood by Bohr and others to necessarily impose
a non-real and non-local representation of the physical world. This
representation is often referred to as the probabilistic interpretation of
quantum mechanics (PIQM) and consists of those postulates thought to
rationally express the physical implications of the underlying compact quantum formalism.

The predictive validity of the quantum formalism is widely acknowledged.
Nevertheless, the PIQM departures from local realism are not universally
accepted. Lepore and Selleri \cite{lepore} contend that \textquotedblleft The
development of local realism since the 1935 Einstein, Podolsky, and Rosen's
paper \cite{einstein}\ is by far the most profound criticism of quantum
mechanics...\textquotedblright\ in reference to the probabilistic
interpretation. Popper raises compelling philosophical arguments in
questioning the validity of PIQM.\cite{popper}

The notable non-local properties of PIQM are manifestations of entanglement
and wave-particle duality. Any viable local alternative to PIQM must minimally
demonstrate an alternative self-consistent basis for both of these phenomena.
In this regard, one of the present authors proposed a locally real
representation of quantum mechanics that gives agreement with performed
experiments for correlated photons and particles \cite{mirell} and is not
restricted by Bell's theorem.\cite{bell} The other phenomenon, wave-particle
duality, is examined by the authors in ref. \cite{mirell2}. In this report we
again examine wave-particle duality and propose a locally real representation
of quantum mechanics, identified here as LRQM, that retains the accepted
underlying quantum formalism with minimal modification.

As a preliminary matter, we first consider some elementary aspects of photon
phenomena from the particular perspective of LRQM as distinguished from PIQM.
In the discrete regime, a single photon is a physically real wave packet
structure on which a real energy quantum resides. The LRQM wave function
$\Phi$ characterizes the wave amplitude of this structure but is not itself a
physical representation of any resident energy quantum. The separability of
these entities is identified with de Broglie's initial representations of
quantum phenomena \cite{debroglie} and is intrinsic to many locally real representations.

Born interpreted the squared modulus of the wave function evaluated at
particular coordinates as a measure of relative probability of finding a
particle on a wave packet.\cite{born} This initial restricted version of
\textquotedblleft Born's rule\textquotedblright, applied here to
electromagnetic radiation for LRQM, is critical in its treatment of the
squared modulus $\left\vert \Phi\right\vert ^{2}$ as a relative and not an
absolute probability flux density. Saxon reminds us of the fundamental utility
of the unnormalized $\left\vert \Phi\right\vert ^{2}$ as the sufficient
determining factor of relative positional probability.\cite{saxon}\ The
subsequent, commonly accepted interpretation of Born's rule incorporates
normalization of the wave function giving unit absolute probability for the
squared modulus integrated over the entire wave packet. Normalization is a
seemingly natural and rational modification of the restricted version and was
fully consistent with the then developing formulation of PIQM. This
modification provides PIQM with a linkage of particle-like and wave-like
properties that compactly expresses both in a wave function $\Phi_{PI}$ but
necessitates the characteristic interpretations of duality and entanglement
for particular phenomena.

Certainly, for an ordinary photon the resident particle-like energy quantum
has unit absolute probability of existing somewhere on the wave packet and
normalization of the LRQM wave function $\Phi$ in this case serves as a
mathematical convenience in equating a unit-valued expression of energy
conservation and a unit-valued probability represented by the integrated
square modulus $\left\vert \Phi\right\vert ^{2}$. Nevertheless, in LRQM an
initial normalization of $\Phi$ must not be arbitrarily re-applied to an
evolving wave function since processes such as interference occurring during
that evolution may create or annihilate wave amplitude. Effectively, LRQM
separates a purely wave-like $\Phi$ from a PIQM $\Phi_{PI}$ and provides that
$\Phi$ with the degree of freedom to scale independent of resident quanta. We
continue to use the term \textquotedblleft probability\textquotedblright\ here
in LRQM bearing in mind that its usage is potentially misleading since that
term suggests equivalence to a mathematical absolute probability, an
equivalence manifested in PIQM as duality.

Classically, the probability flux density $\left\vert \Phi\right\vert ^{2}$ is
recognized as a wave intensity. For a discrete photon incident on an idealized
beam splitter, the wave intensity fractionally divides onto the output
channels in accordance with the transmission and reflection coefficients of
the beam splitter. In the present example, we choose for convenience a 50:50
beam splitter for which the output packets are similar to the incident packet
but with half the intensity giving each a relative probability of $P=0.5$ when
the incident packet is assigned $P=1$. Then, when the energy quantum on that
incident packet reaches the beam splitter, the quantum randomly transfers onto
either of the emerging outgoing $P=0.5$ packets with equal probability. In the
discrete regime this implies that for each incident photon, one of the outputs
is an \textquotedblleft empty\textquotedblright\ wave that is totally
\textquotedblleft depleted\textquotedblright\ in energy quanta relative to its
wave packet probability $P=0.5$. Conversely, the other output is
\textquotedblleft enriched\textquotedblright\ in energy quanta relative to
probability in the regard that the single energy quantum resides on a wave
packet that is now $P=0.5$ instead of $P=1$. The prediction of empty waves in
the discrete regime has been the subject of many investigations seeking to
differentially test local realism and PIQM as in a series of papers by Croca
et al. \cite{croca} as well as in numerous others such as refs. \cite{emp}.
These investigations, frequently using a beam splitter to generate an empty
wave, are necessarily restricted to the very weak wave intensities associated
with discrete photon beams.

As we move from the discrete photon regime to a multiple photon beam in the
continuous wave (cw) regime, the relevant total wave function is
conventionally constructed from a summation over amplitudes of the constituent
wave packets. (Our particular interest here is mono-energetic coherent beams.)
The wave function, which we continue to identify as $\Phi$, is used to express
the beam's intensity $\left\vert \Phi\right\vert ^{2}$, again a probability
flux density. Integration of $\left\vert \Phi\right\vert ^{2}$ over some
selected beam segment gives an inclusive probability $P$. The corresponding
inclusive quanta in that segment have a total energy $E$. We can set $P=E=1$
in arbitrary units for the beam segment.

When this beam is incident on a 50:50 beam splitter, we again have $P=0.5$ on
each of the output channels. However, random statistical distribution of the
inclusive quanta onto the output channels yields $E$ vanishingly close to
$0.5$ on both channels as the inclusive number of quanta becomes large. An
immediate consequence of the cw regime for the beam splitter is that
testability for empty waves is no longer feasible. Then, although local
realism does not expressly prohibit empty waves in the cw regime, the inherent
statistical distribution process is intuitively expected to restrict all
mechanisms for generating empty waves to the discrete regime where conclusive
experimental verification is marginalized.

A mechanism that alters the proportionality of probability and energy on a
photon beam in the cw regime is seemingly as improbable as Maxwell's
hypothesized mechanism for selectively sorting particles based upon their
respective kinetic energies.\cite{maxwell}

In Section III. we present the theoretical basis for a mechanism that, from
the perspective of LRQM, is predicted to selectively alter the proportionality
of probability and energy. In Section IV. we report on the experimental
realization of this phenomenon.

\section{Background}

The theoretical basis developed here for the LRQM wave-like component is
substantially classical. This basis does not obviate the underlying quantum
mechanical formalism such as the wave function associated with quantization of
the electromagnetic field, but it does modify the scaling of a separable,
purely wave-like component $\Phi$. As a consequence, the PIQM duality
proportionality of wave probability (based on that $\Phi$) and particle-like
energy quantum is potentially violated. A mechanism for this duality
violation, trivially but transiently realized in the discrete regime for LRQM,
is however not immediately obvious in the cw regime. Before beginning the
examination of mechanisms that achieve this \textquotedblleft duality
modulation\textquotedblright\ in the cw regime, we return to the example of a
beam splitter once again in order to develop some basic operational
definitions and principles relevant to LRQM.

When a photon is incident on a beam splitter, the transfer process of the
energy quantum to one of the relative probability wave packets on the output
channels is itself of fundamental significance to LRQM. For illustrative
purposes we again choose a 50:50 beam splitter. The quantum on an incident
$P=1$ wave packet enters a zone at the face of the beam splitter where the
quantum randomly transfers onto one of the two emergent $P=0.5$ output wave
packets. Unlike their treatment in PIQM, those output wave packets remain as
real entities at that probability value on both output channels and one packet
does not undergo a collapse when a measurement is made on the other packet.

From the perspective of LRQM, the transfer at the beam splitter must be
treated purely as that of the energy quantum. Wave packet probability
distribution is a non-quantized deterministic process at a device such as the
beam splitter. The energy quantum arriving at the beam splitter surface
randomly transfers onto one of the two emergent outputs causally mediated by
their relative probability distribution consistent with Born's
rule.\cite{born} The wave structures of the two output packets are otherwise
unaltered by the random presence or absence of the quantum on a particular
output packet. (Conversely, we will subsequently consider full photon
transfers in other contexts that are consistent with both PIQM and LRQM, i.e.
duality is not violated. Such transfers may involve scattering conditions that
alter the trajectories of incident wave packets which are then accompanied by
the proportionate energy quanta that had resided on the incident packets.)

In the interests of formalizing the proportionality between wave packet
probability and energy quantum, we assign a value to their proportionality.
Each photon on a discrete beam is represented by an \textquotedblleft
occupation\textquotedblright\ value $\Omega=E/P$ defined as the ratio of the
energy quantum and the total probability of its wave packet.\cite{mirell2} As
a baseline reference, we consider \textquotedblleft ordinary\textquotedblright%
\ photons that might be generated by common atomic emission processes. We
assign $E=1$ to the energy quantum present on a wave packet of probability
$P=1$ giving $\Omega=1/1=1$ for these ordinary photons in arbitrary
dimensionless units.

If a discrete beam composed of such ordinary photons is incident on the 50:50
beam splitter, the output packets have $\Omega=0/0.5=0$ and $\Omega=1/0.5=2$.
Any $\Omega<1$ signifies that the outgoing wave packet is reduced in energy
relative to its associated wave packet probability and is referred to as
\textquotedblleft depleted\textquotedblright. In the extreme case of
$\Omega=0$, the wave packet is totally depleted and is appropriately referred
to as an empty\ wave.\cite{emp} For $\Omega>1$, the wave packet is said to be
\textquotedblleft enriched\textquotedblright. In the present case of
$\Omega=2$, the single quantum resides on a $P=0.5$ wave packet compared to
the $P=1$ wave packet of the incident photon.

In principle, interference of either outgoing wave packet $P=0.5$ with an
independent photon beam would result in the same visibility, unaffected by the
presence or absence of the quantum on that outgoing wave packet.
Fundamentally, a measurement such as interference visibility assesses a beam's
wave-like property whereas a direct detector measurement assesses a beam's
particle-like (energy quantum) property.

As a matter of practical consideration, measurements of the wave-like property
in the discrete regime, as provided for example by beam splitter outputs, are
experimentally problematic. \cite{croca} Accordingly, we are motivated to seek
a duality violation in the cw regime where experimental verification is
significantly enhanced.

In the cw regime, the irradiance $I$ is used to describe the particle-like
component of the photon beam where, conventionally, the formal units are those
of an energy flux density. In contrast, the corresponding wave-like component
is identified as the intensity $W=\left\vert \Phi\right\vert ^{2}$ which, in
the LRQM context, is exclusive of the energy content of the wave. In analogy
to our use of arbitrary units for the discrete beam, we are free to set unit
values for $I$ and $W$ on an ordinary incident cw beam. Any value
specification of flux densities such as $I$ and $W$ implicitly refers to a
sample point on the beam and typically that sample point gives the maxima of
those values e.g. at the beam centroid for a Gaussian cross-section.

With a specified $I$ and $W$ on an incident beam we have $\Omega=I/W$.
Equilibration of energy quanta ensures that the $\Omega$ proportionality is
maintained throughout the beam. Then%
\begin{align}
\Omega &  =\frac{I}{W}\nonumber\\
&  =\frac{\int I(\mathbf{r},t)dadt}{\int W(\mathbf{r},t)dadt}\nonumber\\
&  =\frac{E}{P}%
\end{align}
where the areal integration is over the beam's cross-section and the
integrands are the respective coordinate-dependent values of $I$ and $W$. An
inclusive energy $E$ and probability $P$ on a selected beam segment may be
selected by a choice of temporal integration limits spanning some selected
$\Delta t$. For that $\Delta t$, we are free to assign arbitrary units to
these quantities on the incident beam that give $E=P=1$ whereby $\Omega=1$.
Clearly, from the perspective of LRQM, $\Omega=1$ is maintained on the outputs
where, for a 50:50 beam splitter, $P=0.5$ and, because of random statistical
equilibration, $E=0.5$.

As a result of the common equivalence of the $I/W$ and $E/P$ ratios on any
given beam in the cw regime, we will have frequent occasion to consider either
the $I,W$ (maximum flux density) quantities or the $E,P$ (integrated)
quantities as a matter of illustrative convenience. Within the context of the
present analysis, when $E,P$ is the more convenient pair, the energy flux
(power) $\Delta E/\Delta t$ and the probability flux $\Delta P/\Delta t$ would
be equally convenient. However, we choose $E,P$ in the interests of continuity
with our earlier discussion of discrete photons.%

\begin{figure}
[ptb]
\begin{center}
\includegraphics[
natheight=3.343400in,
natwidth=4.046500in,
height=3.3434in,
width=4.0465in
]%
{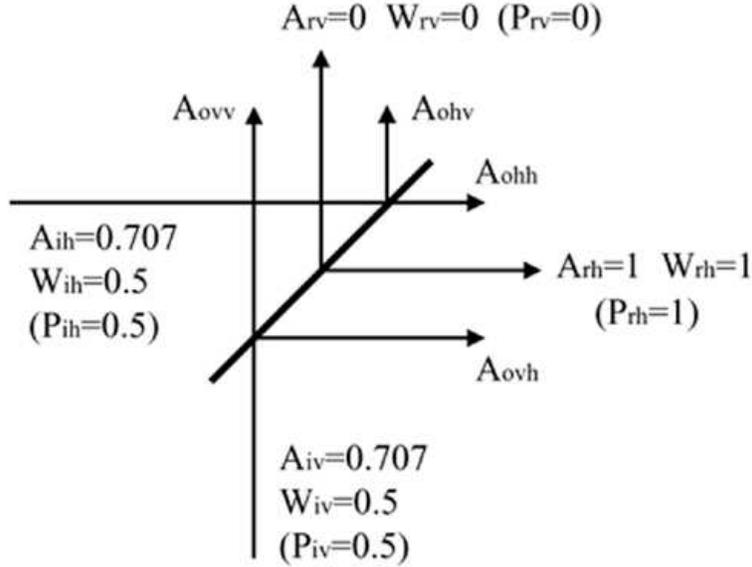}%
\caption{Amplitude moduli associated with two identical mutually coherent
beams coincident on a beam splitter depicted with lateral displacement for
illustrative clarity. Outputs, that form resultants, continue propagating as
real entities. For destructive phasing of $A_{ovv}$ and $A_{ohv}$, the two
incident beams are fully expressed by the resultant $A_{rh}$ as an intensity
output $W_{rh}$. Arbitrary numerical values are assigned for comparative
purposes.}%
\label{f-cap-01}%
\end{center}
\end{figure}
As a reference baseline, an incident beam consisting of photons generated from
conventional sources is identified as ordinary with $\Omega=1$ by appropriate
assignment of the arbitrary energy and probability units. Then the
identifications of a cw beam as depleted or enriched are equivalently
expressed by a respective $\Omega<1$ or $\Omega>1$.

We consider one last example involving the beam splitter, again 50:50, in
which both input channels have mutually coherent similar incident cw beams
with intensities%
\begin{equation}
W_{ih}=W_{iv}=0.5
\end{equation}
directed at the same point on the beam splitter and aligned so as to generate
spatially coincident outputs as identified in Fig. 1. In terms of the
respective probabilities,%
\begin{equation}
P_{ih}=P_{iv}=0.5.
\end{equation}

In this current example, we continue the convention of designating beam
quantities as \textquotedblleft incident\textquotedblright\ prior to
interaction with a mechanism such as the beam splitter and \textquotedblleft
output\textquotedblright\ for the prompt post-interaction manifestations of
those quantities. These post-interaction output quantities then generate
physically distinguishable quantities by interference that are designated as
\textquotedblleft resultant\textquotedblright. Clearly this distinction of
output and resultant designations was superfluous for the previous example of
a single beam incident on a beam splitter. Nevertheless, the utility of these
designation distinctions will be evident when we consider duality modulating mechanisms.

In the present case, the two incident beams intersect at a common point on a
beam splitter. In Fig. 1, the beams are laterally displaced from that common
point in order to clearly depict the various output and resultant components
emanating from that point. The amplitude moduli of the two reflected and the
two transmitted output beams all have identical values%
\begin{equation}
A_{ohv}=A_{ovh}=A_{ohh}=A_{ovv}=0.5
\end{equation}
with corresponding common intensity values%
\begin{equation}
W_{ohv}=W_{ovh}=W_{ohh}=W_{ovv}=0.25
\end{equation}
or, equivalently in terms of respective probabilities,%
\begin{equation}
P_{ohv}=P_{ovh}=P_{ohh}=P_{ovv}=0.25.
\end{equation}
Then for total probabilities in the transition from incidence to output,
\begin{equation}
P_{i}=P_{o}=1,
\end{equation}
and probability is conserved in this incident$\rightarrow$output transition.

Interference between the output beam amplitudes generates a pair of resultant
beams with moduli dependent upon relative incident beam phasing. For our
purposes here, we choose a particular phasing that gives amplitude moduli%
\begin{equation}
A_{rh}=1
\end{equation}
and%
\begin{equation}
A_{rv}=0
\end{equation}
with wave intensities%
\begin{equation}
W_{rh}=1
\end{equation}
and%
\begin{equation}
W_{rv}=0.
\end{equation}

The output pair $A_{ohh}$ and $A_{ovh}$ is fully contiguous to its associated
resultant $A_{rh}$, sharing a common origin at the beam splitter surface. The
same applies to $A_{ovv}$, $A_{ohv}$ and $A_{rv}$. In both cases, the members
of either output pair are not separately experimentally measurable because
interference of those members at their common origin promptly yields the
experimentally measurable resultant wave. Nevertheless, these considerations
do not alter the continued reality of the output wave structures as they
propagate contiguous to and are physically expressed by their respective
resultant waves.

Integration yields the associated probabilities%
\begin{equation}
P_{rh}=\int W_{rh}(\mathbf{r},t)dadt=1
\end{equation}
and%
\begin{equation}
P_{rv}=\int W_{rv}(\mathbf{r},t)dadt=0
\end{equation}
where again, by appropriate choice of integration limits in arbitrary units,
each has the same numerical value as that of the corresponding (maximum) wave
intensity. With the total resultant probability%
\begin{equation}
P_{r}=P_{rh}+P_{rv}=P_{o}=P_{i}=1,
\end{equation}
we have conservation of probability in the output$\rightarrow$resultant
transition as well as in the incident$\rightarrow$output transition.

For our particular choice of relative phase, $P_{rv}$ is identically zero and
$P_{rh}=1$. (That choice could also have been generalized as any relative
phase but in anticipation of a relevant analogy in the next section, we choose
a phase that gives a null value for one of the resultants.) For $P_{rv}$, the
constituent output waves continue to propagate as real but $\pi$ out-of-phase
entities with a null sum. Similarly, the constituent in-phase output waves
generate a resultant $P_{rh}=1$ that exceeds the sum of those output
probabilities,%
\begin{equation}
P_{ohh}+P_{ovh}=0.5.
\end{equation}

These observations are very elementary and would have been superfluous for the
absolute mathematical probabilities of PIQM. However, in the context of the
real wave structures of LRQM, there is a significant process that should be
emphasized here. Interference, which nullifies the resultant of the two output
probabilities on one channel, fortuitously \textquotedblleft
amplifies\textquotedblright\ the output sum probability on the other channel
to precisely compensate for that nullification and conserve probability in the
output$\rightarrow$resultant transition. The criticality of interference in
conserving probability on these real wave structures suggests that a
disruption of this precise compensation by interference may provide a
mechanism for duality modulation.

Then with $\Omega$'s computed from either the $I,W$ or the $E,P$ pairs,%
\begin{equation}
\Omega_{rh}=\Omega_{rv}=\Omega_{r}=1
\end{equation}
respectively for the horizontal, vertical, and total resultant beam occupation
values. These LRQM values are fully consistent with PIQM duality.

The various beam splitter phenomena treated here from the perspective of LRQM
are, nevertheless, almost universally represented in the literature by a
formalism consistent with PIQM. However, in the next section we analogously
apply LRQM to particular grating systems and demonstrate the basis for a
duality-violating mechanism that is not representable by PIQM.

\section{Theoretical basis for duality violation}

\subsection{Gratings with dense sampling}

The general mechanism we consider here for modulating the ratio $\Omega$ of
beam energy and probability is a conventional transmissive \textquotedblleft
picket-fence\textquotedblright\ grating consisting of a regular linear array
of parallel opaque bands with a periodicity $p$ forming intervening slits of
width $w$. Typically for optical gratings of this type, the opaque bands are
thin metallic depositions on one side of a transmissive substrate. In our
analysis, we treat that side as the exit face of the transmission grating with
an incident beam normal to the opposite face of the substrate.

As we proceed, we must take care to adequately quantify the real entities of
energy and probability as incidence$\rightarrow$output and output$\rightarrow
$resultant transitions occur. Ordinary incident beams generated by
conventional sources may be assigned%
\begin{equation}
P_{i}=E_{i}%
\end{equation}
in arbitrary units which gives an incident beam occupation value%
\begin{equation}
\Omega_{i}=1
\end{equation}
as a calculation convenience. The quantification at each transition requires
either a verification that $\Omega$ is maintained or an individual assessment
of energy and of probability in that transition.

For any individual slit, the output energy quanta and the output probability
necessarily remain in constant proportionality relative to that of the
incident beam since that single slit equally samples both quantities from the
incident beam. Indeed, a deviation from this proportionality would constitute
a violation of PIQM duality for a mechanism consisting of a single slit.
Accordingly, from the perspective of LRQM as well as PIQM, an incident
$\Omega_{i}=1$ is maintained on the output of any single slit as well as
collectively over all slit outputs, i.e. $\Omega_{o}=1$. See Fig. 2.

Moreover, with an incident beam of some particular Gaussian diameter $D$, the
(transmitted) output probability and output energy quanta are both invariant
as $w$ and $p$ are proportionately varied since%
\begin{equation}
w/p\equiv\sigma
\end{equation}
presents a constant fractional cross section for transferring probability and
energy quanta from incidence to output, i.e. the grating's transmission
factor. Essentially, proportionate increases in $w$ and $p$ results in
increases for individual slit outputs of both probability and
\begin{figure}
[ptb]
\begin{center}
\includegraphics[
natheight=1.328400in,
natwidth=4.046500in,
height=1.3284in,
width=4.0465in
]%
{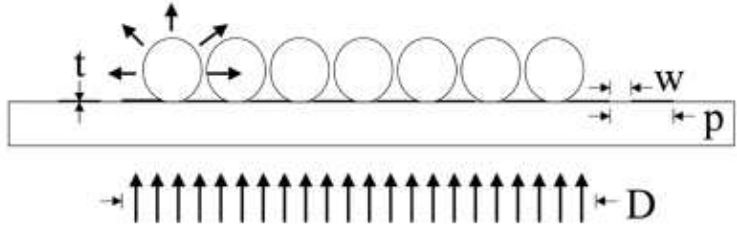}%
\caption{Emergent output diffractive envelopes from $N$ irradiated individual
slits fully determine collective output probability (and energy) prior to the
formation of resultant orders by interference as those envelopes intersect.}%
\label{f-cap-02}%
\end{center}
\end{figure}
energy quanta but, concurrently, the number of incident-irradiated slits
yielding those outputs is proportionately decreased. Then, beginning with an
ordinary incident beam, equal sampling gives the output quantities%
\begin{equation}
P_{o;\sigma}=E_{o;\sigma}=1,
\end{equation}
where both are specific to some particular value of $\sigma=w/p$ but
independent of $w$ and, as a matter of convenience, in arbitrary units we
assign unit value to both in Eq. (20). The output occupation value $\Omega
_{o}$ is constructed from the ratio of $P_{o;\sigma}$ and $E_{o;\sigma}$ but,
because of equal sampling for those quantities for any $\sigma$, their ratios%
\begin{equation}
\Omega_{o}=\Omega_{i}=1
\end{equation}
are independent of $\sigma$.

The transition to the resultant diffraction orders requires that we examine
quantities such as wave amplitude and intensity for the emergent output
diffraction envelopes. In this regard it is most expedient to begin with the
classic phasor construction for the output wave intensity from a single slit
$W_{os}$ from Kirchhoff diffraction theory in the Fraunhofer approximation%
\begin{equation}
W_{os}(\alpha)=W_{os}(0)\frac{\sin^{2}\alpha}{\alpha^{2}}=W_{os}%
(0)sinc^{2}\;\alpha
\end{equation}
where%
\begin{equation}
\alpha=\frac{\pi w}{\lambda}\sin\theta
\end{equation}
for a wavelength $\lambda$. Eq. (22) above must be used with care as we
proceed since the intensity $W_{os}(\alpha)$ is expressed in \textquotedblleft%
$\alpha$-space\textquotedblright\ rather than over the physical $\theta$
angular space azimuthal to the slit. Moreover, this phasor construction
relates entirely to the slit output intensity distribution, giving
$W_{os}(\alpha)$ relative to a centroid value $W_{os}(0)$. Effectively then,
the Eq. (22) intensity is unscaled with respect to the incident beam intensity
attenuation in transiting the slit of width $w$. We will return to this
consideration below and show that it is readily resolved. In the meantime, as
a matter of convenience, we choose arbitrary units with $W_{os}(0)=1$.

Our present investigation also does not consider slits in the sub-wavelength
range $w<\lambda$. Nevertheless, we will have need to consider slit widths in
the neighborhood of $w\gtrsim\lambda$ where classic references such as that of
Elmore and Heald caution us that \textquotedblleft the Kirchhoff diffraction
theory is less accurate, and it is expected that the single-slit diffraction
factor will no longer give a good description of the
envelope.\textquotedblright\cite{elmore} This caution is quite correct.
However, we will return to this point and show that such deviations are not a
factor in the essential criteria for duality violation. Accordingly, we
proceed here with the use of the $sinc^{2}\alpha$ envelope in the interests of
illustrating duality violation using an explicit analytical function for that
envelope and, later in this section, we show that the essential criteria for
duality violation are well satisfied in the present case and identify the
general requirements of an arbitrary envelope in meeting these criteria.

From Eq. (22), we can write the expression for the output single slit
probability%
\begin{equation}
P_{os}(\alpha_{t})=\int_{-\alpha_{t}}^{\alpha_{t}}sinc^{2}\;\alpha\;d\alpha
\end{equation}
where envelope truncation%
\begin{equation}
\alpha_{t}=\frac{\pi w}{\lambda}%
\end{equation}%
\begin{figure}
[ptb]
\begin{center}
\includegraphics[
natheight=3.636500in,
natwidth=4.046500in,
height=3.6365in,
width=4.0465in
]%
{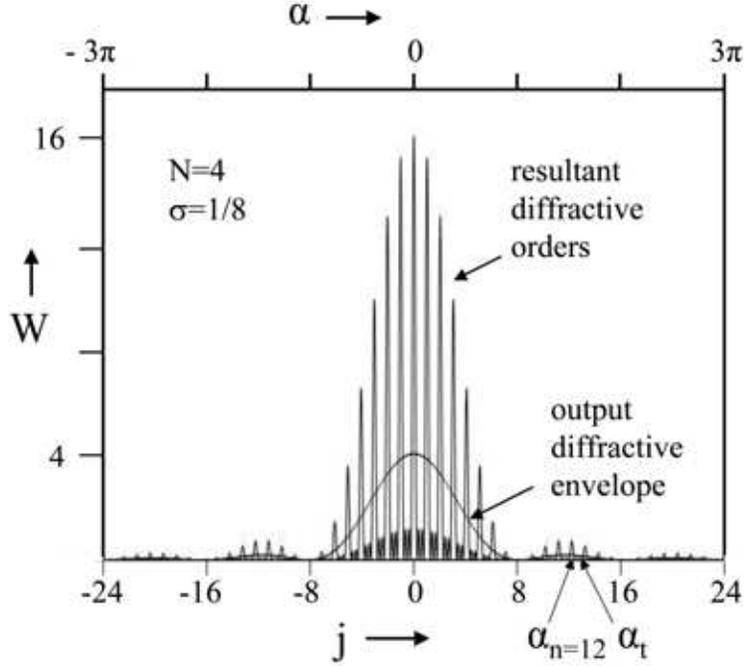}%
\caption{Intensity plots in arbitrary units of the single-slit output
diffractive envelope $sinc^{2}\alpha$ scaled by a factor $N$, the number of
irradiated slits, to give a smoothly varying intensity envelope in $\alpha
$-space representing the collective output probability of the grating, and the
diffractive order peaks representing the resultant probability. An untypically
small$\ N=4$ is chosen here for clarity of depiction. The integrals of the two
intensity functions are essentially equal, demonstrating probability
conservation for the $\sigma=1/8$ approximation of dense sampling.}%
\label{f-cap-03}%
\end{center}
\end{figure}
found from Eq. (23) with $\theta=90%
{{}^\circ}%
$ provides the appropriate limits of the Eq. (24) integral. In analogy to the
collective output probability of two beams incident on a beam splitter, the
collective output probability of the grating is%
\begin{equation}
P_{o;\sigma}(\alpha_{t})=N\int_{-\alpha_{t}}^{\alpha_{t}}sinc^{2}%
\;\alpha\;d\alpha
\end{equation}
where here, relative to $P_{os}(\alpha_{t})$, the coefficient $N$ is
effectively the number of slits irradiated by the incident beam. $N$ is well
determined for a given beam incident beam diameter $D$, $\sigma$, and
$\alpha_{t}$. The $Nsinc^{2}\alpha$ intensity envelope is shown in Fig. 3. The
Eq. (24) and Eq. (26) probabilities are identified as functions of the
integration limit $\alpha_{t}$. From Eq. (25) we see that those probabilities
are equivalently functions of $w$. Fig. 3 shows an example of an $\alpha_{t}$
for a narrow slit $w\gtrsim\lambda$. The truncation of the $sinc^{2}\alpha$
envelope by $\pm\alpha_{t}$ at this $w$ results in a central lobe and partial
side lobes as output from each slit with substantial envelope distribution
from $+90%
{{}^\circ}%
$ to $-90%
{{}^\circ}%
$. Conversely, significantly increasing $\alpha_{t}$ results in a wide
$w\gg\lambda$, adds multiple, greatly diminished side lobes to each slit
output and concurrently confines the significant envelope contribution to the
forward direction clustered about $0%
{{}^\circ}%
$.

However, the Eq. (24) probability $P_{o;\sigma}(\alpha_{t})$ computed in
$\alpha$-space, is clearly an increasing function of $\alpha_{t}$, or
equivalently, of $w$. This result is in conflict with the Eq. (20)
$w$-independent probability $P_{o;\sigma}$ deduced from physical
considerations for gratings of varying $w$ but having some specific $\sigma$.
The conflict can be understood by noting that the Eq. (24) dependence on $w$
is an artifactual consequence of integrating the unscaled Eq. (22) intensity
in $\alpha$-space rather than in physical space over the $\theta$ angle
azimuthal to the slit. This can be appreciated by examining the limit of large
$w$ for which the $w$-dependence of $P_{o;\sigma}(\alpha_{t})$ becomes
vanishingly small. In this limit, the significant contributions to the
integral are entirely confined to small $\alpha$ where $\alpha=(\pi
w/\lambda)\sin\theta\rightarrow\pi w\theta/\lambda$ and $\alpha$ in this range
is linear in $\theta$. Nevertheless, the Eq. (26) collective output
probability will prove to be extremely useful precisely because of its
$w$-dependence in $\alpha$-space.

We next turn to the interference-generated resultant probability emerging from
the intersecting individual slit output $sinc$ $\alpha$ amplitude envelopes
(shown in Fig. 2) that generate the output $sinc^{2}\alpha$ intensity
envelopes. With the incident beam of some Gaussian diameter $D$ spanning a
large number N of grating slits, the non-null resultant probability computed
in $\alpha$-space is effectively confined to the sum of the individual
integrals of the narrow, highly directional principal order intensity peaks,
and the intermediate subsidiary peaks are vanishingly small.

The classic Kirchhoff expression for the resultant diffraction order intensity
is given by%
\begin{equation}
W_{r}(\alpha)=W_{os}(0)sinc^{2}\alpha\left(  \frac{\sin(N\alpha/\sigma)}%
{\sin(\alpha/\sigma)}\right)  ^{2}%
\end{equation}
where $W_{os}(0)=1$ is, as before, the single slit intensity at $\alpha=0$ set
to unity. The quantity $(\sin(N\alpha/\sigma)/\sin(\alpha/\sigma))^{2}$ is the
\textquotedblleft grating factor\textquotedblright. In this form, the
diffraction peak maxima extend above the single slit $sinc^{2}\alpha$
intensity envelope but are proportional to it.

In the literature, Eq. (27) is generally modified by replacing $W_{os}(0)$
with the corresponding collective $N$ slit intensity $W_{o}(0)$ at $\alpha=0$
and inserting a compensating $N^{-2}$ factor based upon the justification that
the $N$ slit intensity at $\alpha=0$, $W_{o}(0)=N^{2}W_{os}(0)$. That
modification would yield the familiar depiction of a diffraction envelope with
the maxima of the diffraction order peaks in contact with that envelope.
However, we deliberately omit that modification here in the interests of
correctly tracking the relative probability in the output$\rightarrow
$resultant transition. Output quantities appropriately refer to
pre-interference tabulations of those quantities. If the single slit output
intensity is $W_{os}(0)$ at $\alpha=0$, the collective output intensity from
$N$ slits is $NW_{os}(0)$. The same principle is involved with the Eq. (24)
single slit output probability and the Eq. (26) collective output probability.

Consequently, the integral of the Eq. (22) single slit output scaled by $N$,
i.e. $NW_{os}(\alpha)$, is correctly equal to the integrated Eq. (27) as
depicted in Fig. 3. The computed equality of these integrals for any selected
limits $\pm\alpha_{t}$ is a confirmation that probability is conserved in the
output$\rightarrow$resultant transition. Exact integral equality is achieved
as $\sigma\rightarrow0$ for which the sample density of the output envelope by
the resultant orders becomes infinite. However, even the small but finite
$\sigma=w/p=1/8$ selected in Fig. 3 for illustrative clarity, gives an
approximation of \textquotedblleft dense sampling\textquotedblright. In this
approximation, integral equality is closely achieved as a function of the
integration limits $\pm\alpha_{t}$ with only minor perturbations as those
limits pass between discrete resultant orders such as the depicted $\alpha
_{t}$ displaced from the $n=12$ order.

Again, for purposes of illustrative clarity in Fig. 3, an $N=4$ has been
selected, which is lower than that for typical experimental conditions by
about two orders of magnitude, so that the envelope and the diffraction order
peaks can both be represented on the same scale. Low intensity secondary
diffraction order peaks appearing near the bases of the principal diffraction
order peaks in the figure diminish to negligible values for experimentally
realistic large $N$.

The integral of the diffraction envelope, which gives the collective output
probability in $\alpha$-space, has limits $\pm\alpha_{t}$ where the envelope
truncates at the physical azimuthal angles $\theta=\pm90%
{{}^\circ}%
$ relative to the grating normal. These limits are a function of $w$ by
$\alpha_{t}=w\pi/\lambda$. The narrow resultant peaks (principal maxima) arise
from interference of the intersecting single-slit diffraction envelope
amplitudes. The peaks occur at $\alpha=\alpha_{j}=j\pi\sigma$ which is
$j\pi/8$ in the figure. The resultant probability can be expressed as a
summation in $\alpha$-space with limits $\pm\alpha_{n}$ where $\left\vert
\alpha_{n}\right\vert \lesssim\left\vert \alpha_{t}\right\vert $. For small
$w$, the grating is \textquotedblleft fine\textquotedblright, the limits in
$\alpha$-space are low and $w\gtrsim\lambda$. For $j=12$, $\alpha_{12}=3\pi/2$
and a grating with a truncation value marginally inclusive of the $\pm12^{th}$
orders, i.e. $\alpha_{t}\gtrsim\alpha_{12}$, has $w\gtrsim3\lambda/2$.
Conversely, for large $w$ (not shown in the figure), the grating is
\textquotedblleft coarse\textquotedblright, the limits in $\alpha$-space are
high and $w\gg\lambda$.%

\begin{figure}
[ptb]
\begin{center}
\includegraphics[
natheight=3.230900in,
natwidth=4.046500in,
height=3.2309in,
width=4.0465in
]%
{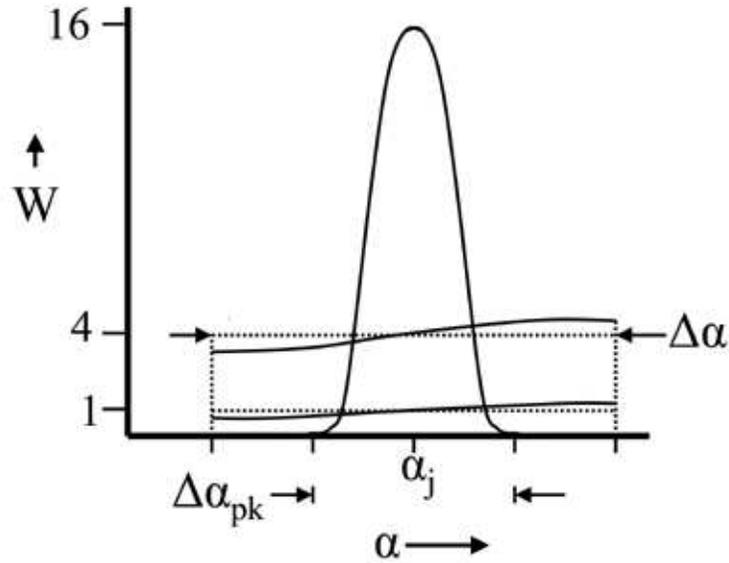}%
\caption{ A detail of a single subinterval of Fig. 3 with $N=4$ and
$\sigma=1/8$ showing some arbitrary $j^{th}$ peak, the associated single slit
output intensity envelope and the collective (four) slit output intensity
envelope with its associated Riemann approximation. The width of the
subinterval is $\Delta\alpha=\pi\sigma$ and the width of the peak is
$\Delta\alpha_{pk}=\pi\sigma/2$ for the present $N=4$. Most critically, the
figure depicts the probability equivalence of the peak and the collective
output intensity envelope over the subinterval.}%
\label{f-cap-04}%
\end{center}
\end{figure}
Probability conservation in the output$\rightarrow$resultant transition can
also be assessed for individual resultant peaks. From Eq. (27), the grating
factor separates the interference maxima into subintervals%
\begin{equation}
\Delta\alpha=\pi\sigma
\end{equation}
and, within those subintervals, the null to null peak base width is%
\begin{equation}
\Delta\alpha_{pk}=2\pi\sigma/N.
\end{equation}
A detail of a subinterval is shown in Fig. 4. The total output probability for
this subinterval is%
\begin{align}
P_{oj;\sigma}(\alpha_{j})  &  =N\int_{\alpha_{j}-\Delta\alpha/2}^{\alpha
_{j}+\Delta\alpha/2}sinc^{2}\alpha\;d\alpha\nonumber\\
&  =N\pi\sigma\;sinc^{2}\alpha_{j}%
\end{align}
where the area under the single slit $sinc^{2}\alpha$ curve at $\alpha_{j}$ is
approximated by the product of $\Delta\alpha=\pi\sigma$ and the functional
value $sinc^{2}\alpha_{j}$. (The value of $\alpha$ at the $j^{th}$ order is
$\alpha_{j}=j\pi\sigma$ where the applicable $\sigma$ is implicit from the
context.) The total subinterval output probability is given by that area,
$\pi\sigma sinc^{2}\alpha_{j}$, multiplied by the number $N$ of contributing
slits. We are reminded again that this factor giving total output probability
is linear in $N$ since it relates to an accounting of the collective single
slit probability outputs prior to the subsequent interference that yields the
resultant probabilities. Total output probability in the earlier example of
two beams incident on a beam splitter provides a useful analog. As a result of
the linear factor $N$ in Eq. (30), the collective output intensity envelope is
appropriately depicted as the single slit output intensity envelope
$sinc^{2}\alpha_{j}$ necessarily scaled up by a factor $N$ as shown in Fig. 4.
This scaling still leaves the resultant diffraction peak maxima above the
collective output envelope but now graphically illustrates the essential
conservation of probability in the output$\rightarrow$resultant transition for
this single resultant peak.

Conservation of probability can also be demonstrated quantitatively for this
arbitrary single $j^{th}$ resultant peak shown in Fig. 4. The probability for
this peak is found from Eq. (27) noting that%
\[
\frac{\sin\left(  N\alpha_{j}/\sigma\right)  }{\sin(\alpha_{j}/\sigma)}=N
\]
at the principal maxima $\alpha=\alpha_{j}$. Then%

\begin{align}
P_{rj;\sigma}(\alpha_{j})  &  =\int_{\alpha_{j}-\Delta\alpha_{pk}/2}%
^{\alpha_{j}+\Delta\alpha_{pk}/2}sinc^{2}\alpha\;\left(  \frac{\sin\left(
N\alpha/\sigma\right)  }{\sin(\alpha/\sigma)}\right)  ^{2}d\alpha\nonumber\\
&  =\left(  \frac{1}{2}\right)  \left(  \frac{2\pi\sigma}{N}\right)  \left(
N^{2}sinc^{2}\alpha_{j}\right) \nonumber\\
&  =N\pi\sigma\;sinc^{2}\alpha_{j}%
\end{align}
where the integral of the $j^{th}$ peak is readily evaluated by noting that,
because of peak symmetry about $\alpha_{j}\pm\pi\sigma/2N$, the peak occupies
one-half of the rectangular area defined by the peak null to null base width
$\Delta\alpha_{pk}=2\pi\sigma/N$ and magnitude $N^{2}sinc^{2}\alpha_{j}$.
Those factors resolve to a resultant subinterval probability and an output
subinterval probability, both of which can be equated to a
Riemann-approximated subinterval area, i.e.%
\begin{equation}
P_{rj;\sigma}(\alpha_{j})=P_{oj;\sigma}(\alpha_{j})=\pi\sigma(Nsinc^{2}%
\alpha_{j}).
\end{equation}
Accordingly, over all $\alpha$-space we can define the total resultant
probability as a Riemann sum%
\begin{equation}
P_{r;\sigma}(\alpha_{n})=\sum_{j=-n}^{n}\pi\sigma(Nsinc^{2}\alpha_{j})
\end{equation}
and the total output probability as the corresponding definite integral%
\begin{equation}
P_{o}(\alpha_{t})=\int_{-\alpha_{t}}^{\alpha_{t}}Nsinc^{2}\alpha_{j}\;d\alpha
\end{equation}
where $\Delta\alpha=\pi\sigma\rightarrow d\alpha$. We have limit equivalence
$\alpha_{n}=\alpha_{t}$ in Eqs. (33-34) when $\alpha_{t}=n\pi\sigma$. We can
temporarily defer consideration of deviations of the continuous-valued
$\alpha_{t}$ variable from the discrete-valued $\alpha_{n}$ variable since we
are currently imposing dense sampling in $\alpha$-space. Dense sampling
divides the integration partition into a large number of subintervals ($n\gg
1$). Consequently, $\alpha_{t}>\alpha_{n}$ results in fractional integration
of Eq. (34) into the $\pm(n+1)$ subintervals not included in the Eq. (33)
summation. The contribution of those two fractionally integrated subintervals
relative to that of the total $2n+1$ subintervals partition is vanishingly small.

As a formal matter, in the dense sampling limit $\sigma\rightarrow0$, the
partition subinterval $\Delta\alpha\rightarrow0$ which gives $n\rightarrow
\infty$ and%
\begin{equation}
\lim_{n\rightarrow\infty}\sum_{j=-n}^{n}\Delta\alpha\;sinc^{2}\;\alpha
_{j}=\int_{-\alpha_{t}}^{\alpha_{t}}sinc^{2}\;\alpha\;d\alpha
\end{equation}
where the integration limits $\pm\alpha_{t}$ are applicable since $\alpha
_{n}\rightarrow\alpha_{t}$. The limit condition in Eq. (35) is recognized as
the Riemann sum equivalency to the definite integral on the right. Ultimately,
the Riemann sum serves as a mathematical intermediary that demonstrates
output$\rightarrow$resultant probability equivalency when $\alpha
_{n}\rightarrow\alpha_{t}$.

Nevertheless, despite their equivalency, the $\alpha$-space quantities
$P_{o}(\alpha_{t})$ and $P_{r;\sigma}(\alpha_{n})$ are not satisfactory
expressions of probabilities since an examination of output probability for
some fixed $\sigma$ predicts $w$-independence. Both $P_{o}(\alpha_{t})$ and
$P_{r;\sigma}(\alpha_{n})$ are clearly increasing functions of $w$ through
their respective $w$-dependent respective limits $\alpha_{t}$ and $\alpha_{n}%
$. However, because of the relative equality of $P_{o}(\alpha_{t})$ and
$P_{r;\sigma}(\alpha_{n})$, the $w$-dependence of $P_{r;\sigma}(\alpha_{n})$
can be removed by normalization \ with $P_{o}(\alpha_{t})$ which has the same
$w$-dependence. The normalized resultant probability%
\begin{align}
P_{r;\sigma}(\alpha_{t})  &  =\frac{\sum_{j=-n}^{n}\pi\sigma\;sinc^{2}%
\alpha_{j}}{\int_{-\alpha_{t}}^{\alpha_{t}}sinc^{2}\alpha_{j}\;d\alpha
}\nonumber\\
&  =1
\end{align}
gives a $w$-independent constant to within a vanishingly small discrepancy
arising from $\alpha_{t}>\alpha_{n}$ as discussed above.

The summation limits $\pm n$ are set by the condition $\left\vert \alpha
_{n}\right\vert \leq\left\vert \alpha_{t}\right\vert $. Note that the
identification of the Eq. (36) normalized expression $P_{r;\sigma}(\alpha
_{t})$ differs from that of the Eq. (33) unnormalized expression $P_{r;\sigma
}(\alpha_{n})$ only with regard to the $\alpha$-space limits. The summation in
$P_{r;\sigma}(\alpha_{t})$ has the same $\pm n$ limits as that of
$P_{r;\sigma}(\alpha_{n})$ but the more distal integration limits $\left\vert
\alpha_{t}\right\vert \geq\left\vert \alpha_{n}\right\vert $ of the Eq. (34)
normalization integral sets $\alpha_{t}$ as the defining functional variable
of Eq. (36). Because of dense sampling, Eq. (36) is very nearly a constant
and, as such, is effectively independent of $\alpha_{t}$. However, as a
formality we retain $\alpha_{t}$ as an apparent functional dependent. (Since
$\alpha_{t}=\pi w/\lambda$, $\alpha_{t}$-invariance is equivalent to
$w$-invariance for Eq. (36) where $\lambda$ is a constant.)

From a purely pragmatic viewpoint, the normalization in Eq. (36) expeditiously
achieves the required $w$-invariance of $P_{r;\sigma}(\alpha_{t})$ for the
class of gratings with dense sampling. More significantly, however, the Eq.
(36) normalization of relative resultant probability by relative output
probability (distinct from PIQM normalization, Section I) is fundamental in
quantifying any probability non-conservation in the output$\rightarrow
$resultant transition. For separate real entities of energy quanta and
probability, the constant proportionality required by PIQM duality is no
longer a constraint. In LRQM, deviations from duality may occur as
non-conservation of probability manifested as a change in the ratio of the
resultant probability relative to the output probability. Therefore, for a
mechanism that potentially achieves duality violation by probability
non-conservation in some transition process such as grating output$\rightarrow
$resultant, it is critical to represent that resultant probability relative to
the base output probability in order to express that non-conservation.

Non-conservation of probability is, however, not in evidence as $\sigma
\rightarrow0$ provides dense sampling of the outputs by the resultants. Then
the Eq. (36) normalized form $P_{r;\sigma}(\alpha_{t})$ rectifies the
artifactual increase of the Eq. (33) $P_{r;\sigma}(\alpha_{n})$ with $w$ and
provides the physical probability as a function of $\alpha_{t}$, albeit a
trivially constant unit value.

In analogy to the beam splitter with two incident beams, we again have
conservation of probability in the output$\rightarrow$resultant transition and
we also have spatial regions of null resultant probability and regions of
\textquotedblleft amplified\textquotedblright\ resultant probability.
Similarly, the manifestation of these resultant probabilities does not alter
the reality of the output probability wave structures. The grating, however,
differs from the beam splitter in that the origin of the resultants is
physically displaced from the grating surface but still in the near field
where the emergent expanding individual slit output envelopes begin to
intersect and interfere as shown in Fig. 2.

To complete the analysis of the grating with dense sampling, we need to also
assess energy in the output$\rightarrow$resultant transition. As the expanding
output probability envelopes interfere, the resident energy quanta transfer
without loss onto the forming resultant diffraction order beams. The process
is purely an energy quanta transfer. No wave entity is transferred in a
process analogous to that of energy quanta transfer at a beam splitter. The
consequent continued proportionality of total energy and probability in the
output$\rightarrow$resultant transition yields a constant $\Omega$ throughout,%
\begin{equation}
\Omega_{i}=\frac{E_{i}}{P_{i}}=\Omega_{o}=\frac{E_{o}}{P_{o}}=\Omega_{r}%
=\frac{E_{r}}{P_{r}}=1.
\end{equation}
We note that in the Eq. (36) $P_{r;\sigma}(\alpha_{t})$ expression, the
property of dense sampling is critical to maintaining an $\Omega_{r}=1$
independent of $w$. As $\alpha_{t}$ ($=\pi w/\lambda$) is increased, the
normalization integral monotonically increases whereas the summation abruptly
increases with two additional final terms $\pm n^{\prime}=\pm\left\vert
n+1\right\vert $ as $\alpha_{t}$ reaches a value corresponding to a new
$j=n^{\prime}$ order. Nevertheless, for dense sampling, the additional $\pm
n^{\prime}$ terms are incrementally small relative to the existing sum of the
$2n+1$ terms..

\subsection{Gratings with sparse sampling}

The dependence upon dense sampling for the maintenance of $\Omega\approx1$
suggests that the converse, i.e. \textquotedblleft sparse\textquotedblright%
\ sampling, constitutes a potential mechanism for achieving significant
deviations of $\Omega$ from unity which is equivalently a violation of PIQM
duality. Specifically, we examine $\sigma=0.5$, characteristic of Ronchi
rulings. This ratio $\sigma$ is notable in that the side lobes of the output
probability are each entirely expressed by the single, odd order that
bifurcates those lobes in $\alpha$-space as shown in Fig. 5. This is a
significant fortuitous property of a $\sigma=0.5$ grating attributable to the
non-expression by the null-valued even orders on either end of each of those
lobes. It reflects a corollary of Eq. (36) that it is necessarily the non-null
resultants that express the outputs. The probability for $\sigma=0.5$,%
\begin{equation}
P_{r;0.5}(\alpha_{t})=\frac{(\pi/2)\sum_{j=-n}^{n}sinc^{2}\;\alpha_{j}}%
{\int_{-\alpha_{t}}^{\alpha_{t}}sinc^{2}\;\alpha\;d\alpha},
\end{equation}
is merely a special case of the Eq. (36) probability where we again consider
$w\gtrsim\lambda$ (fine gratings) to $w\gg\lambda$ (coarse gratings).
Similarly,%
\begin{equation}
P_{rj;0.5}(\alpha_{t})=\frac{(\pi/2)sinc^{2}\;\alpha_{j}}{\int_{-\alpha_{t}%
}^{\alpha_{t}}sinc^{2}\;\alpha\;d\alpha}%
\end{equation}
for the $j^{th}$ order probability.%

\begin{figure}
[ptb]
\begin{center}
\includegraphics[
natheight=2.335900in,
natwidth=4.046500in,
height=2.3359in,
width=4.0465in
]%
{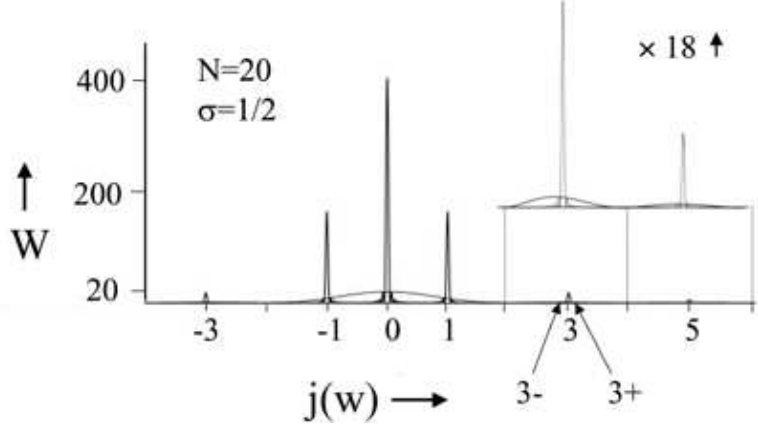}%
\caption{A $\sigma=0.5$ sparse sampling analog to the dense sampling
approximation shown in Fig. 3. Odd $j$ peaks for $j\geq3$ have probabilities
that fully express the probability of the collective output lobe that they
respectively bifurcate.}%
\label{f-cap-05}%
\end{center}
\end{figure}
However, for the sparse sampling associated with $\sigma=0.5$,%
\begin{equation}
\alpha_{j}=j\pi/2
\end{equation}
and the Eq. (38) probability, unlike the Eq. (36) probability with
$\sigma\rightarrow0$, is no longer a constant of the functional variable
$\alpha_{t}$. The deviations from constancy occur as $\alpha_{t}$ approaches
the neighborhood of the odd orders. Eq. (38) is plotted in Fig. 6 in
$\alpha_{t}$-space from $\alpha_{t}=\pi$ to $3\pi$. For $\alpha_{t}=\pi$,
truncation occurs at the $\pm2^{nd}$ orders, representing a relatively fine
grating, whereas $\alpha_{t}=3\pi$ corresponds to truncation at the $\pm
6^{th}$ orders and the grating is modestly coarser. The horizontal axis on the
figure shows only positive values since symmetry ensures that any choice of
$\alpha_{t}$ automatically sets integration limits of $\pm\alpha_{t}$ and
concurrently sets summation limits $\pm n$ in the evaluation of $P_{r;0.5}%
(\alpha_{t})$.

An examination of Fig. 6 notably shows a probability edge transition of
$\sim5\%$ for $\alpha_{t}$ in the neighborhood of the $\pm3^{rd}$ orders. The
transitions at the odd orders diminish in magnitude as $\alpha_{t}$ increases
further, ultimately yielding flat-line constancy of $P_{r;0.5}(\alpha_{t})$
for very large $\alpha_{t}$ (coarse) gratings.%

\begin{figure}
[ptb]
\begin{center}
\includegraphics[
natheight=2.840000in,
natwidth=4.046500in,
height=2.84in,
width=4.0465in
]%
{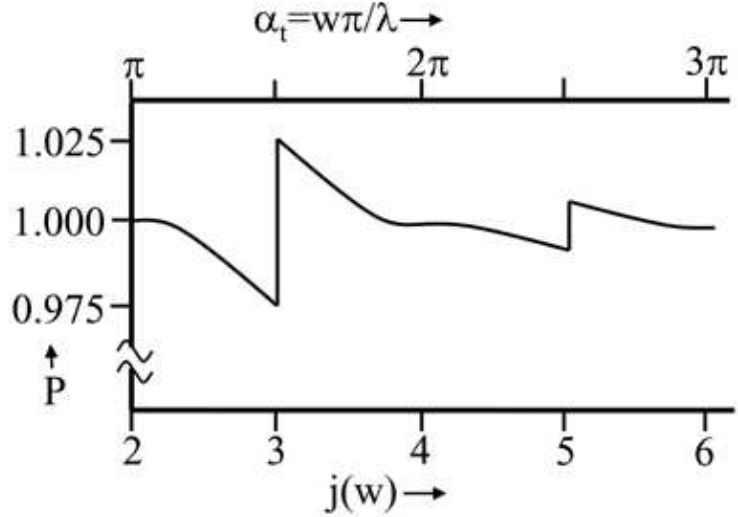}%
\caption{Resultant probability $P_{r;0.5}(\alpha_{t})$ plotted here for
positive $\alpha_{t}$ since functional evaluation of the probability at limits
$\alpha_{t}=\pm w\pi/\lambda$ is understood. The excursions from unity
represent measurable non-conservation of probability. These excursions are
notable for a grating that has a $w$ for which $\alpha_{t}$ is in the
neighborhood of $\alpha_{j=3}$.}%
\label{f-cap-06}%
\end{center}
\end{figure}
The functional structure of the Fig. 6 probability is, of course, simply a
consequence of the sparse sampling of the $\sigma=0.5$ gratings. As
$\alpha_{t}$ at $\alpha_{j=2}$ ($\pm2^{nd}$ order) is increased, the
$P_{r;0.5}(\alpha_{t})$ summation is constant whereas the normalization
integral increases resulting in a maximum $\sim2.5\%$ decrease in the limit as
$\alpha_{t}$ approaches $\alpha_{j=3}$ from the left. At $\alpha_{j=3}$,
$P_{r;0.5}(\alpha_{3})$ discontinuously increases by $\sim5\%$ with the
inclusion of the $j=\pm3^{rd}$ terms in the summation. As $\alpha_{t}$ is
increased toward $\alpha_{j=4}$, the summation with end terms at $j=\pm
n=\pm3$, is again constant while the normalization integral continues to
increase resulting in a return to the probability value at $j=2$. This
functional behavior repeats with increasing $\alpha_{t}$ but with
progressively diminishing discontinuities at the odd $j$. (Since the resultant
beams are physically narrow but finite in Gaussian diameter, $P_{r;0.5}%
(\alpha_{t})$ only approximates a true mathematical discontinuity at odd $j$.)
The convergence of $P_{r;0.5}(\alpha_{t})$ to a constant, unit value in the
limit as $\alpha_{t}\rightarrow\infty$ verifies that the Eq. (38) probability
is properly scaled.

Then upon inspection of Fig. 6, we conclude that probability is annihilated in
the output$\rightarrow$resultant transition for $\sigma=0.5$ gratings with
$\alpha_{t}\lesssim\alpha_{j}$ at odd $j$. Conversely, probability is created
in the output$\rightarrow$resultant transition for $\sigma=0.5$ gratings with
$\alpha_{t}\gtrsim\alpha_{j}$ at odd $j$.

As in the above case of $\alpha_{t}$ and $\alpha_{j}$, we have applied a
convention in which truncation-related quantities are associated with a
resultant $j^{th}$ diffraction order. It will frequently be convenient to
return to this convention. For our purposes here, this association is made
with a positive-valued $j$ but by symmetry the association also applies to the
corresponding negative-valued $j$ (with, of course, a reversal in the above inequalities).%

\begin{figure}
[ptb]
\begin{center}
\includegraphics[
natheight=2.222600in,
natwidth=4.046500in,
height=2.2226in,
width=4.0465in
]%
{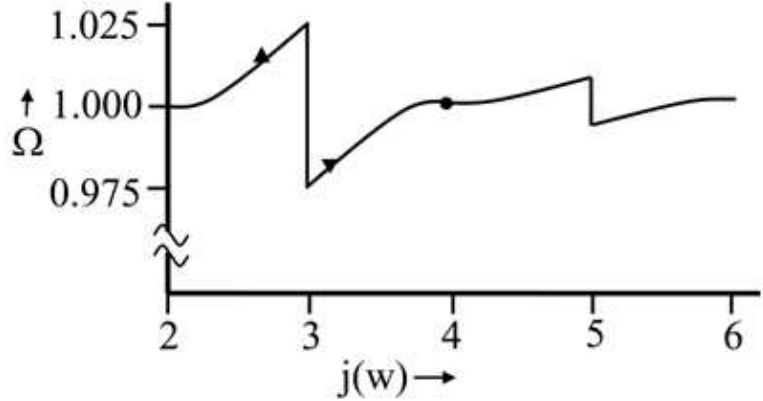}%
\caption{Theoretical occupation value $\Omega_{r;0.5}(\alpha_{t}%
)=\Omega_{G[j(w)]\text{ }th}$ for $\sigma=0.5$. With energy $E$ set to unit
value, this function is the inverse of that in Fig. 6. The regions above the
unity level identify grating $w$-values that give enrichment whereas those
regions below unity correspondingly are associated with depletion. The three
experimentally measured $\Omega^{\prime}s$ are plotted for comparison on this
theoretical curve at their respective $j$-equivalent values of truncation:
$\Omega_{G(2.63)\text{ }ex}$ as $\blacktriangle$, $\Omega_{G(3.16)\text{ }ex}$
as $\blacktriangledown$, and $\Omega_{G(3.94)\text{ }ex}$ as $\bullet$.}%
\label{f-cap-07}%
\end{center}
\end{figure}
We proceed with an examination of the Eq. (38) probability by computing the
associated occupation value for this class of gratings,%
\begin{align}
\Omega_{r;0.5}(\alpha_{t})  &  =\frac{E_{r;0.5}(\alpha_{t})}{P_{r;0.5}%
(\alpha_{t})}\nonumber\\
&  =\frac{\int_{-\alpha_{t}}^{\alpha_{t}}sinc^{2}\;\alpha\;d\alpha}%
{(\pi/2)\sum_{j=-n}^{n}sinc^{2}\;\alpha_{j}}%
\end{align}
since the output energy, which can be set to unity, is conserved in the
transition to resultant energy, $E_{o}=1=E_{r;0.5}$. Functionally,
$\Omega_{r;0.5}(\alpha_{t})$ plotted in Fig. 7, is the inverse of the Fig. 6
probability $P_{r;0.5}(\alpha_{t})$. $\Omega_{r;0.5}(\alpha_{t})$ then also
converges to unity as $\alpha_{t}\rightarrow\infty$. For $\sigma=0.5$ gratings
with $\alpha_{t}\lesssim\alpha_{j}$ at odd $j$, we have enrichment of the
resultant diffraction orders. Conversely, for $\sigma=0.5$ gratings with
$\alpha_{t}\gtrsim\alpha_{j}$ at odd $j$, those orders are depleted.

It is instructive at this juncture to detail the respective diffraction order
parameters $P$, $E$, and $\Omega$ for a pair of Ronchi gratings identified as
$G(3\pm)$ that have truncation on either side of the significant $\pm3^{rd}$
orders. Care should be taken to avoid confusion of these two $\pm$
designations. For $G(3-)$, the positive-valued $\alpha$-space truncation is
denoted as $\alpha_{3-}=\alpha_{t}\lessapprox\alpha_{3}$, which marginally
excludes the formation of the $\pm3^{rd}$ order resultant probability channels
by locating $+\alpha_{t}$ at the \textquotedblleft interior\textquotedblright%
\ (left) null at the base of the $+3^{rd}$ order peak. By symmetry, the
negative-valued $-\alpha_{t}$ is situated at the interior (right) null of the
$-3^{rd}$ order peak. Similarly, for $G(3+)$ the positive-valued $\alpha
$-space truncation is denoted as $\alpha_{3+}=\alpha_{t}\gtrapprox\alpha_{3}$,
which marginally allows the formation of the $\pm3^{rd}$ order resultant
probability channels at threshold by situating $\pm\alpha_{t}$ at the
respective exterior nulls of the $\pm3^{rd}$ order peaks. For a given
operating wavelength $\lambda$, the two Ronchi gratings $G(3\pm)$ are
characterized entirely by their respective slit widths $w_{3\pm}$. From Eqs.
(25) and (40) at the $+3^{rd}$ order threshold the requisite width is
$w_{3}=3\lambda/2$. Therefore the gratings $G(3-)$ and $G(3+)$ have respective
widths $w_{3-}\lessapprox3\lambda/2$ and $w_{3+}\gtrapprox3\lambda/2$.

Note that in the above quantities, we have extended the convention of
referencing truncation to a $j^{th}$ order by use of $j-$ to indicate marginal
truncation exclusion of the $\pm j^{th}$ orders and $j+$ to indicate marginal
truncation inclusion of the $\pm j^{th}$ orders.%

\begin{figure}
[ptb]
\begin{center}
\includegraphics[
natheight=2.335900in,
natwidth=4.046500in,
height=2.3359in,
width=4.0465in
]%
{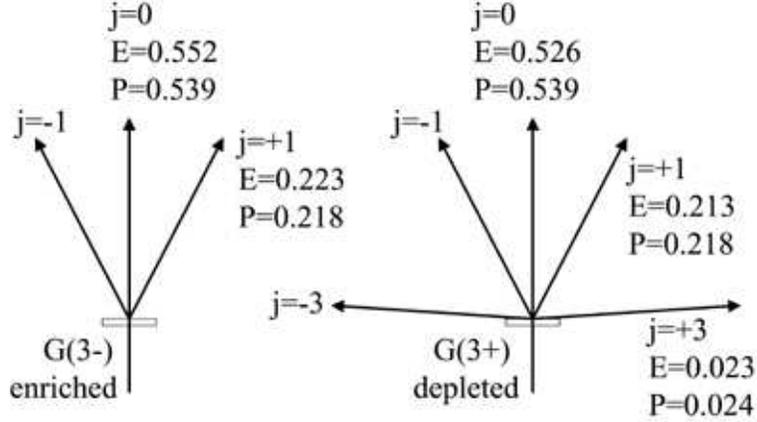}%
\caption{Tabulation of energy and probability values on the resultant
propagating orders for gratings $G(3-)$ and $G(3+)$ where the total energy is
unit valued. By symmetry, values on $\pm j$ are the same for both gratings.}%
\label{f-cap-08}%
\end{center}
\end{figure}
The $P$, $E$, and $\Omega$ parameters for these two gratings $G(3\pm)$ are
shown in Fig. 8 where the collective and the individual $j^{th}$ probabilities
are computed from Eqs. (38) and (39), respectively. The output energy
$E_{o}=1$ is conserved giving a total of $E_{r}=1$ distributed onto the
propagating orders as $E_{rj}$. Consistent with probability as a relative
quantity, these $E_{rj}$ are found from the ratio of the Eq. (39) individual
$j^{th}$ order resultant probability $P_{rj;0.5}(\alpha_{t})$ and the Eq. (38)
collective resultant probability $P_{r;0.5}(\alpha_{t})$. The Fig. 8\ example
illustrates the essential features of probability non-conservation and energy
distribution in the neighborhood of the $\pm3^{rd}$ order threshold
probability discontinuity.

We note that since this discontinuity alters the proportionality of the
wave-like probability and the particle-like energy, the phenomenon can be
expressed as a \textquotedblleft duality modulation\textquotedblright. In the
Fig. 8 example, the $G(3-)$ grating with $\pm\alpha_{3-}$ truncation points
provides a $+2.5\%$ duality modulation upon comparing the $G(3-)$ occupation
value $\Omega_{G(3-)}$ to the ordinary value $\Omega=1$. Similarly, $G(3+)$
provides a $-2.5\%$ duality modulation.

Most generally, this phenomenon of duality modulation is potentially
applicable to any (fine) grating that yields a small number of resultant
orders for which one or two can be located near or just beyond the plane of
the grating \textquotedblleft at threshold\textquotedblright. In other words,
the precise shape of the output diffraction envelope is not a critical factor.
This can be understood by the relationship of the diffraction envelope and the
interference factor that modulates that envelope. Within any single
subinterval, the $j^{th}$ peak arises from the same interference factor,
identified earlier in Eq. (31), modulating a value $f(\alpha_{j})$ of an
output diffraction envelope $f(\alpha)$ in place of $sinc^{2}\alpha$.
Similarly, in the present context of the regular Ronchi grating where
$w\gtrsim\lambda$, the departure from the Fraunhofer approximation
$w\gg\lambda$ gives a single slit output intensity envelope that is not
precisely expressed by $sinc^{2}\alpha$. However, orders near threshold for
gratings with sparse sampling are still predicted to give probability non-conservation.

It may be appreciated at this point that the existence of expressed resultant
probability discontinuities for fine, sparsely sampled gratings could have
been deduced directly and quite succinctly from basic LRQM principles. The
essential rationale for such a deduction is derived from the observation that,
for a set of gratings with a common sparse sampling, e.g. $\sigma\sim0.5$, all
such gratings have the same output probability, but we can always identify
specific pairs of these gratings with nearly identical slit widths $w_{j+}$
and $w_{j-}$ for which $w_{j+}$ marginally admits a significant resultant
order near threshold and $w_{j-}$ marginally blocks that order. For these two
gratings considered in succession, the collective resultant probability must
exhibit an abrupt decrease in the near infinitesimal $w_{j+}\rightarrow
w_{j-}$ transition. That abrupt decrease does not occur for the smoothly
varying output probability. Similarly, the output energy is also unaltered by
that transition. Therefore, in the absence of an energy-dissipative mechanism,
the distribution of the output energy onto the resultants yields a relative
enrichment of the $w_{j+}$ resultants with respect to the $w_{j-}$ resultants.
However, without the basis developed in the previous subsection for densely
sampled gratings, the assessment of occupancy in threshold transitions is
limited to relative changes in occupancy. Moreover, the occupancy is not
characterized outside the neighborhood of the $w_{j}$ threshold value.

Accordingly, we have proceeded here in a less succinct manner by first
examining elementary beam splitter configurations to elucidate the basic
principles of probability conservation from the perspective of LRQM. We then
examined resultant probability associated with gratings having dense sampling
in order to develop a generalized functional expression applicable to all
gratings that characterizes resultant probability over the full $\alpha$-space
and not just in the neighborhood of threshold transitions before proceeding to
an examination of gratings having sparse sampling.

It is of some peripheral interest that the spatial physical discreteness of
resultant diffraction orders of a grating is the essential property that
produces duality modulation in the present example. We recall that it was also
spatial physical discreteness that yielded duality modulation for the case of
a discrete photon beam incident on a beam splitter. For a grating, it is the
discrete increment of probability (on a highly directional diffraction order)
that modulates duality whereas for the beam splitter, it is the discrete
increment of energy that produces this modulation.

We conclude this section on the theoretical basis for duality violation with a
brief but relevant examination of grating anomalies. This examination is
necessitated by some apparent similarities of those anomalies to the phenomena
considered here.

Grating anomalies, since their discovery by Wood in 1902, have continued to be
a subject of intensive experimental and theoretical investigation.\cite{wood}
These anomalies are deviations, often abrupt, in the diffraction order
irradiances as a function of a parameter such as wavelength. These deviations
were initially designated as anomalies since they departed from predictions of
earlier classical principles thought to fully characterize grating phenomena.
Their continued characterization as \textquotedblleft
anomalies\textquotedblright\ is now generally regarded as a misnomer since
sophisticated theoretical electromagnetic analyses of these phenomena have
substantially approximated the experimental observations.

Our particular interest here concerns \textquotedblleft
Rayleigh\textquotedblright\ anomalies that have a theoretical basis postulated
by Lord Rayleigh several years after their discovery by Wood.\cite{rayleigh}%
\ Rayleigh anomalies are characterized by abrupt increases in the irradiance
of propagating orders as one of those orders \textquotedblleft at
threshold\textquotedblright\ near the grating plane is extinguished by an
incremental wavelength increase. The essential theoretical basis postulated by
Lord Rayleigh consists of coherent (photon) scattering of the threshold order
on a grating's periodic structures. By interference, the scattered threshold
order (inclusive of its irradiance) is coherently re-distributed onto the
remaining propagating orders in proportion to the relative intensities of
those orders. This coherent scattering process is fully consistent with PIQM
which was developed two decades later. The postulated theoretical basis
transfers photon energy quanta as well as photon wave probabilities to the
remaining propagating orders thereby maintaining duality.

We seek here an alternative demonstration in LRQM of Rayleigh-like anomalies
in the absence of the coherent scattering process postulated by Lord Rayleigh.
With $\sigma=0.5$, the slit width $w$ is selected as the dependent parameter
consistent with our previous analysis. From Eq. (25), for which $w$ and
$\lambda$ are inversely related, LRQM predicts abrupt irradiance increases as
orders reach threshold with progressive incremental decreases in $w$.

In the open interval between any successive $j=n$ and $j+1$, each of the
individual $2j+1$ propagating orders exhibits a proportionately decreasing
probability across that interval as shown in Eq. (39) and Fig. 6. (As in the
Sec. 3 analysis, we consider the positive orders, but the corresponding
negative orders are inclusive by symmetry for normal incidence.) The resultant
probability of any single order relative to the total resultant probability
Eq. (38) is constant for any $\alpha_{t}$ within any such open interval. This
fractional resultant probability for
\begin{figure}
[ptb]
\begin{center}
\includegraphics[
natheight=3.230900in,
natwidth=4.046500in,
height=3.2309in,
width=4.0465in
]%
{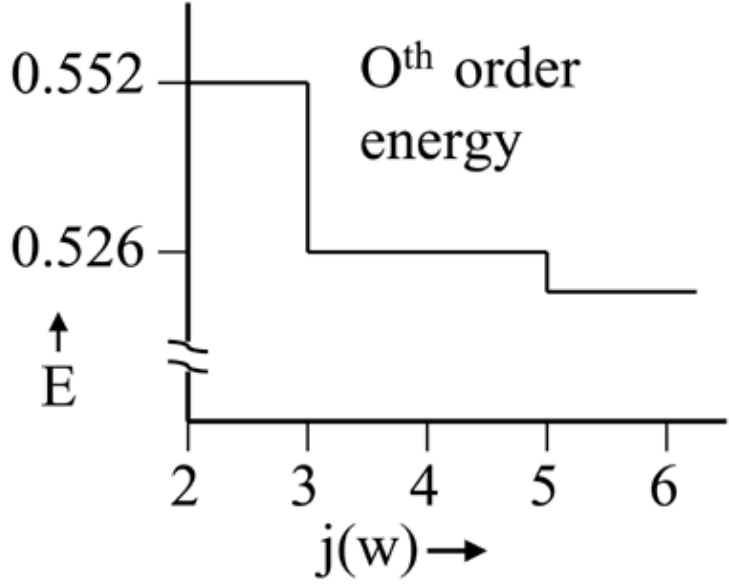}%
\caption{0$^{th}$ order energy as a function of $j(w)$-equivalent truncation
showing characteristic edge drops in propagating energy (or irradiance)
associated with Rayleigh anomalies at threshold values of significant orders.
Compare to Fig. 8.}%
\label{f-cap-09}%
\end{center}
\end{figure}
the $0^{th}$ order%
\begin{align}
R_{0}(\alpha_{t})  &  =\frac{P_{r0;0.5}(\alpha_{t})}{P_{r;0.5}(\alpha_{t}%
)}\nonumber\\
&  =\frac{1}{\sum_{j=-n}^{n}sinc^{2}\;\alpha_{j}}\nonumber\\
&  =R_{0}(\alpha_{n}).
\end{align}
is found from the ratio of Eq. (39) with $j=0$ and Eq. (38). The independence
of this ratio with respect to the $\alpha$-space location $\alpha_{t}$ is
emphasized by that ratio's equivalence to evaluation at the initial $\alpha
$-space location $\alpha_{n}$ of that interval, i.e. $R_{0}(\alpha_{n})$.
Essentially, the $0^{th}$ order's fractional share of the total resultant
probability is calculated with respect to the probability of all propagating
orders as a function of increasing $\alpha_{t}$. This fractional share is a
step function, constant for some total number of propagating orders $2n+1$ and
then discontinuously dropping as $\alpha_{t}$ reaches the next order at some
$j=n+1$. The $0^{th}$ order's fractional share, calculated against the total
probability of the $2(n+1)+1$ propagating orders, drops because of the
additional probability on the $\pm\left\vert n+1\right\vert $ orders. Fig. 9
shows a plot of the $0^{th}$ order fractional share $R_{0}(\alpha_{t})$.

In LRQM, energy quanta entering a probability field equilibrate onto that
field in proportion to relative probabilities. Since the total output energy
quanta in that field are independent of $\alpha_{t}$, the $0^{th}$ order's
fractional share of total resultant probability $R_{0}(\alpha_{t})$ also gives
the fraction of the total output energy $E_{o}$ quanta equilibrating to the
$0^{th}$ order. Then the energy on the $0^{th}$ order is%
\begin{equation}
E_{r0}=E_{o}R_{0}(\alpha_{t})
\end{equation}
and with $E_{o}=1$ in arbitrary units, $E_{r0}$ is simply represented by the
Fig. 9 $R_{0}(\alpha_{t})$ step function. This function predicts the abrupt
step-wise increases associated with Rayleigh anomalies both with regard to
their locations at threshold $\alpha_{t}$ and their magnitudes relative to the
irradiance of the paired extinguished threshold orders.

However, unlike the (PIQM-consistent) coherent scattering process postulated
by Lord Rayleigh, the LRQM-consistent basis described earlier in this section
also predicts a duality violation as quanta transfer from an extinguished
threshold order to the remaining propagating orders without concurrently
transferring wave intensity to those orders.

We stress that this finding does not preclude the existence of PIQM-consistent
postulated coherent scattering as the operant mechanism for Rayleigh anomalies
in other gratings and beam configurations, but it does expand the potential
mechanisms that may be the cause of observed anomalies to include those that
are not consistent with PIQM. In this regard we note that Lord Rayleigh based
his theoretical analysis on experimental evidence from reflective gratings
having deep, pronounced periodic groove structures which would be expected to
exhibit significant scattering of a threshold order into the field of the
remaining propagating orders. Conversely, the transmissive gratings we
consider here consist of periodic flat opaque bands of negligible thickness
that would not provide a comparable scattering mechanism.

\section{Experiment}

\subsection{Apparatus and beam parameters}

In this subsection we begin with a description of the experimental
configuration shown in Fig. 10 before proceeding in subsequent subsections to
the methodology for assessing duality violation on that configuration.%

\begin{figure}
[ptb]
\begin{center}
\includegraphics[
natheight=2.222600in,
natwidth=5.361800in,
height=2.2226in,
width=5.3618in
]%
{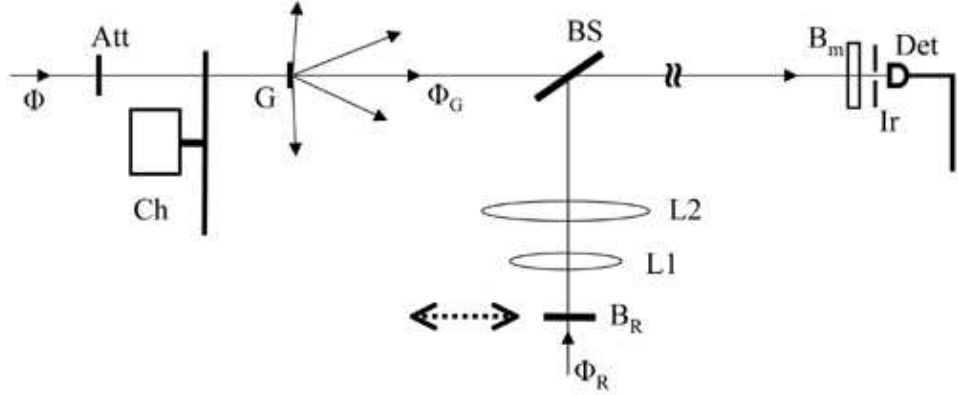}%
\caption{Experimental apparatus configuration\ showing a potentially
duality-modulated beam $\Phi_{G}$ following passage through a particular
grating $G$. $\Phi_{G}$ is coupled to an ordinary beam $\Phi_{R}$ along a
coupling path extending from a beam splitter $BS$. Coupling occurs with beam
blocker $B_{R}$ shifted to transmit $\Phi_{R}$.}%
\label{f-cap-10}%
\end{center}
\end{figure}
A HeNe laser generates a horizontally linearly polarized beam $\Phi$ of
several milliwatts at 633 nm. Beam $\Phi$ traverses a variable attenuator
$Att$ and an optical beam chopper wheel $Ch$. $\Phi$ is normally incident on a
grating $G$. The grating is one of several Ronchi rulings with various grating
slit widths. The intervening opaque bands of the gratings defining those slits
are 150 nm-thick reflective chromium deposited on a glass substrate. The
particular grating under study is mounted with the ruling on the exit face and
the bands vertically oriented. The $0^{th}$ order diffraction beam identified
as $\Phi_{G}$ is incident on a 50:50 beam splitter $BS$. An independent, HeNe
laser generates a horizontally linearly polarized beam $\Phi_{R}$ initially
several milliwatts in power. $\Phi_{R}$ passes a retractable beam blocker
$B_{R}$ and enters a beam expander, $L\,1\ $($f=+100$ $mm$) and $L\,2$
($f=+200$ $mm$), before forming a beamspot concentric with that of $\Phi_{G}$
on the beam splitter $BS$ as shown in the Fig. 11 detail. This concentricity
is a critical alignment for the apparatus. As a practical matter, the
experimental configuration includes a number of beam directors not shown in
Fig. 10 that facilitate beam alignment and folding of optical paths.

As we shall see in the subsequent subsections, measurement of duality
violation imposes some general criteria on the beam \ and apparatus
parameters. However, in the interests of facilitating replication of the
present experiment, we provide specific parameters here in greater detail than
that necessitated by the general criteria.

The (Gaussian) diameter of $\Phi_{G}$ at $BS$ is $\sim2mm$ by natural
divergence from the source laser and whereas the corresponding diameter of
$\Phi_{R}$ at $BS$\ is $\sim4mm$ as a result of $L1$ and $L2$. The beam
components exiting $BS$ utilized here are the transmitted component of
$\Phi_{G}$\ and the reflected component of $\Phi_{R}$. The orientation of
$BS$\ is adjusted to coaxially align the $\Phi_{R}$\ beam spot to the
$\Phi_{G}$\ beam spot at the terminus of a $2000mm$ \textquotedblleft coupling
path\textquotedblright. This critical, second coaxial beamspot alignment
effectively coaxially aligns the $\Phi_{R}$\ and $\Phi_{G}$\ beams on the
coupling path. The variable attenuator $Att$ significantly reduces the beam
power of $\Phi$ such that $\Phi_{G}$ along the coupling path is $\sim6\mu W$.
The corresponding beam power of $\Phi_{R}$ along the coupling path is
approximately two orders of magnitude higher at $\sim600\mu W$. $\Phi_{G}$
expands to a Gaussian diameter of $\sim4.5mm$ at the terminus of the coupling
path by natural divergence from the emitting laser whereas $\Phi_{R}$
converges to $\lesssim1.7mm$ by setting the relative spacing of lenses $L1$
and $L2$.%

\begin{figure}
[ptb]
\begin{center}
\includegraphics[
natheight=1.524700in,
natwidth=5.054800in,
height=1.5247in,
width=5.0548in
]%
{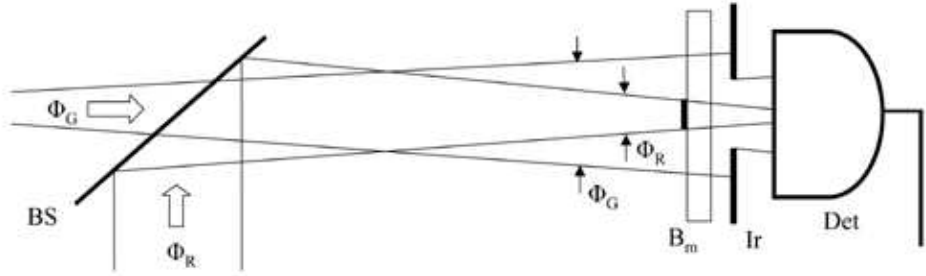}%
\caption{Detail of Fig. 10 coupling path showing the substantial separation of
$\Phi_{R}$ from $\Phi_{G}$ at the path terminus by convergence of the former
onto a disk mask.}%
\label{f-cap-11}%
\end{center}
\end{figure}
An assembly of a beam blocker $B_{m}$, an iris diaphragm $Ir$, and a
photodiode detector $Det$ is located at the coupling path terminus. $B_{m}$
consists of a $1.7mm$ opaque disk mask mounted on a transparent glass
substrate. The iris is set to a $3.3mm$ diameter. Beam directors on the
coupling path (not shown in Figs. 10 and 11) are used to provide concentric
alignment of $\Phi_{G}$ and $\Phi_{R}$ with $B_{m}$ and $Ir$. These settings
collimate the $\sim6\mu W$ $\Phi_{G}$ beam on the coupling path to an annular
beam of $\sim2.4\mu W$ incident on the adjacent detector $Det$. Of the
remaining power, $\sim1.5\mu W$ (25\%) is blocked by the disk mask $B_{m}$ and
$\sim2.1\mu W$ (35\%) is blocked by the diaphragm of $Ir$. The $\sim600\mu W$
power of $\Phi_{R}$ on the coupling path is blocked almost entirely by $B_{m}$
alone leaving a residual $\sim2.5\mu W$ incident on the detector. The function
of the iris setting is to simultaneously restrict the radial sampling of
$\Phi_{G}$ to an annular region closely coupled to $\Phi_{R}$ while
concurrently providing adequate $\Phi_{G}$ power for detector measurement.

\subsection{Observation of duality violation}

We briefly digress in this subsection from the experimental apparatus and
consider the theoretical basis for observing duality violation. As we
discussed in Section III, orders passing to threshold for particular gratings
are associated with abrupt redistributions of irradiance onto the remaining
propagating orders for Rayleigh anomalies. A similar redistribution is
predicted here for particular Ronchi gratings but, additionally, with the
predicted phenomenon of duality violation on those remaining propagating
orders. Consequently, for these Ronchi gratings the duality state of the
propagating orders must be measured in order to establish whether those orders
are ordinary, as would be expected for Rayleigh anomalies, or are in fact
duality modulated. The experimental design used to achieve this measurement
requires a transient coupling between a resultant presumptively
duality-violating beam from the grating (a propagating order) and an
independent ordinary beam.\cite{mirell2}

This design is analogous to the intersecting of two independent beams as used
in numerous investigations \cite{hull}\ to experimentally assess duality
violation by determining the presence or absence of interference between those
beams, i.e. to provide a test of PIQM. A review of these investigations is
given by Paul.\cite{paul} An observation of interference would seemingly
violate Dirac's dictum that a photon (in PIQM) can interfere only with
itself.\cite{dirac}

The outcomes of these numerous investigations are conclusive demonstrations
that interference does occur in apparent contradiction to PIQM. However, in a
theoretical analysis of this phenomenon, Mandel makes the critical argument
that for any given photon measured in the interference we do not know on which
beam that photon had initially resided.\cite{mandel} Because of that lack of
knowledge, each photon is treated in Mandel's analysis as interfering with
itself. Consequently, interference in the intersection of independent beams is
consistent with PIQM and does not provide a test of that interpretation.
Concurrently, that interference is trivially consistent with LRQM.

In a variant of those independent beam investigations, we prepare one of the
beams by transmission through a particular grating where we have some basis
that a prepared beam $\Phi_{G}$ may specifically violate PIQM duality, i.e.
the beam is in a depleted or enriched state from the perspective of LRQM but
is necessarily ordinary for PIQM. Spatially transient equilibration of that
prepared beam with a second, ordinary beam $\Phi_{R}$ by mutual interference
over the coupling path should then yield a net transfer of energy quanta for
LRQM but not for PIQM. That net energy transfer relative to $\Phi_{G}$ is
readily measurable in the cw regime by detecting disparate beam powers on the
sampled $\Phi_{G}$ with and without $\Phi_{R}$ present on the coupling path.

\subsection{Gratings}

The particular gratings used in the experiment are based upon the theoretical
analysis presented in Section III. In the apparatus shown in Fig. 10, the
horizontally linearly polarized ($\lambda=633nm$) $\Phi$ is normally incident
on one of three Ronchi transmission gratings $G$ with respective slit widths
$w$ that are theoretically predicted to generate resultant orders that are
respectively enriched, depleted and ordinary. For whichever of the three
gratings is installed in the Fig. 10 apparatus, the lines on the exit face of
that particular grating are vertically oriented thereby providing TM (S)
polarization with respect to $G$ in the usual classical configuration for
observing grating anomalies.

From Eq. (25), for a specified incident wavelength, the slit width uniquely
determines the truncation point $\alpha_{t}$. From the theoretical analysis in
Section III, the most relevant attribute of a particular Ronchi grating is its
$\alpha_{t}$ truncation point with respect to the location of some particular
$j^{th}$ order. The three Ronchi gratings used in this investigation are
uniquely characterized by the respective slit widths $w=833$ $nm$, $1000$
$nm$, and $1250$ $nm$. From Eqs. (25) and (40), $j(w)=2w/\lambda$ gives a
continuum of $w$-dependent truncation points relative to the $j^{th}$ integer
diffraction orders that is%
\begin{equation}
j(w)=2.63,\text{ }3.16,\text{ and }3.94
\end{equation}
for the three respective slit widths. In the present investigation,
$\lambda=633$ $nm$ is invariant. Accordingly, the individual gratings are also
uniquely characterized by the Eq. (44) continuum $j(w)$-equivalent truncation
points. A\ Ronchi grating of some arbitrary slit width $w$ is denoted as
$G[j(w)]$ or simply $G$. Conversely, a grating identified with a numerical
$j(w)$-equivalent truncation point identifies a particular grating with an
implicitly expressed slit width. This use of $j(w)$ continues the convention
in which variables are most instructively identified by the critical
$j$-equivalent truncation value.

In this convention we are reminded that a grating such as $G(3)$ has
truncations precisely bisecting the $\pm3^{rd}$ orders whereas a grating
$G(3-)$ has truncations marginally exclusive of the $\pm3^{rd}$ orders i.e.
truncations at the respective interior nulls of the $\pm3^{rd}$ peaks.
Extending this to the $j(w)$ continuum, a particular grating $G(2.63)$ has an
implied slit width $w=833$ $nm$. The truncation point at $j=2.63$ is exclusive
of the $\pm3^{rd}$ orders but only approximately satisfies the marginal
exclusion of those orders specified by $j=3-$. Similarly, a grating $G(3.16)$
implies $w=1000$ $nm$ and includes the $\pm3^{rd}$ orders, but the truncation
point at $j=3.16$ is not marginal as it is for $j=3+$. Despite the lack of
truncations marginally close to $j=3$ for $G(2.63)$ and for $G(3.16)$, these
gratings are nevertheless predicted to yield orders that are respectively
significantly enriched and depleted.

The final grating, $G(3.94)$ with $w=1250$ $nm$, is included to provide for a
control experiment since the $j$-equivalent truncation point very closely
matches the null orders at $j=4$. See Fig. 7. Gratings with a $j$-equivalent
truncation in the immediate neighborhood of $j=4$ are predicted to provide
orders that are ordinary. Consequently, $G(3.94)$ orders are expected to have
no significant net energy transfer from coupling and should then exhibit no
duality violation.

\subsection{Calculation methods}

Data are acquired with one of the three Ronchi gratings in the Fig. 10
position of $G$. Since the Gaussian beam diameters of the coupled $\Phi_{G}$
and $\Phi_{R}$ are not equal over the coupling path as a result of lenses
$L1$\ and $L2$, we necessarily undertake the examination of coupling phenomena
with the extensive variables $E$\ and $P$\ rather than their respective
intensive flux densities irradiance $I$ and wave intensity $W=\left\vert
\Phi\right\vert ^{2}$. We continue the use of $\Phi_{G}$ and $\Phi_{R}$ as
general identifiers of the respective beams, but we are reminded that these
wave functions in LRQM are exclusive of the energy quanta residing on those beams.

The basic premise of beam coupling is that a duality modulated beam
equilibrates with an ordinary beam by a net transfer of energy quanta that
leaves the wave structures of both beams unchanged and ideally converges the
occupation values toward a common value. This convergence objective
constitutes criterion (1). For this idealized coupling, a fully equilibrated
state is achieved as $\Phi_{G}$ and $\Phi_{R}$ approach the end of the
coupling path giving the equality%
\begin{equation}
\Omega_{Gc}=\Omega_{Rc}.
\end{equation}
We use the added subscript \textquotedblleft$c$\textquotedblright\ on
variables such as $\Omega$ to denote values at the end of the coupling path
where equilibration of $\Phi_{G}$ and $\Phi_{R}$ has potentially altered those
values relative to their respective values without $\Phi_{G}$ and $\Phi_{R}$
simultaneously present on the coupling path for those same variables.

Under a criterion (2), $\Phi_{R}$ should ideally serve as an infinite source
for a depleted $\Phi_{G}$ or an infinite sink for an enriched $\Phi_{G}$ in
the equilibration process. In the present case with%
\[
P_{R}\gg P_{G}%
\]
satisfied by a respective ratio of $\sim100:1$ for these respective
probabilities, we have an approximation of criterion (2) leaving the final
equilibrated $\Phi_{G}$ and $\Phi_{R}$ both as ordinary and extending the Eq.
(45) $\Omega$ equality to a unit-valued ordinary value,%
\begin{equation}
\Omega_{Gc}=\Omega_{Rc}=1.
\end{equation}

If $\Phi_{G}$ is depleted or enriched as it emerges from a grating $G$, a net
transfer of energy $\Delta E$ will occur between $\Phi_{G}$ and $\Phi_{R}$
that changes the initial grating-emergent energy $E_{G}$ of $\Phi_{G}$ to
\begin{equation}
E_{Gc}=E_{G}\pm\Delta E
\end{equation}
as the coupling path terminus is approached. A positive-signed $\Delta E$
corresponds to an energy gained by $\Phi_{G}$ in a transfer from $\Phi_{R}$
where $\Phi_{G}$ had initially been depleted. Similarly, if $\Phi_{G}$ had
initially been enriched, $\Delta E$ is negatively signed. Alternatively, if
$\Phi_{G}$ emerging from $G$ is initially ordinary, no net transfer occurs and
$\Delta E$ is zero.%

\begin{figure}
[ptb]
\begin{center}
\includegraphics[
natheight=2.537400in,
natwidth=3.044100in,
height=2.5374in,
width=3.0441in
]%
{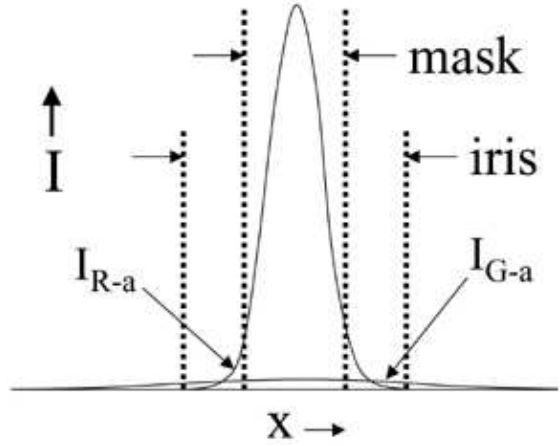}%
\caption{Graphical depiction of irradiances at the coupling path terminus
showing, in particular, the respective $\Phi_{G}$ and $\Phi_{R}$ annular beam
irradiances $I_{G-a}$ and $I_{R-a}$ incident on the detector $Det$.}%
\label{f-cap-12}%
\end{center}
\end{figure}
The coupling equilibration of $\Phi_{G}$ to an ordinary state (if it is not
already in an ordinary state) provides at the coupling path terminus in our
arbitrary units the important result%
\begin{equation}
P_{G}=E_{Gc}.
\end{equation}
A sampling of this $E_{Gc}$ is readily acquired by the detector.

Under idealized criterion (3), $\Phi_{R}$ is entirely excluded from the
detector annular sampling region. In the present Fig. 11 configuration with
the given beam parameters, the $\sim100:1$ ratio of $\Phi_{R}$ probabilities
$P_{R-m}$ blocked by the $B_{m}$ mask and $P_{R-a}$ incident on the detector
sampling region approximates criterion (3) as depicted in the Fig. 12 graph of irradiances.

The Fig. 13 oscilloscope pattern for the detector sampling with the chopper
wheel in motion is a square wave of the pulsed $\Phi_{G}$ energy with a
baseline biased by some steady state energy on $\Phi_{R}$ residually in the
annular sampling field as well as any background level. Then the oscilloscope
peak height measurement of the square wave%
\begin{align}
\Delta V_{Gc}  &  =\kappa E_{Gc}\nonumber\\
&  =\kappa P_{G}%
\end{align}
gives the beam's post-coupled energy and the $\Phi_{G}$ probability as well
because of Eq. (48) to within a multiplicative constant $\kappa$. Chopper
wheel pulsing of the $\Phi_{G}$ beam enables a peak height measurement of the
square wave that automatically separates the detector sampling of $E_{Gc}$
from any steady-state bias sources. For acquisition with a particular Ronchi
grating $G$ in position, the chopper wheel transmits and blocks $\Phi_{G}$ for
equal 5 msec time intervals giving 10 msec cycles in generating a square wave
pulse train transmitted to the oscilloscope from the detector amplifier.%

\begin{figure}
[ptb]
\begin{center}
\includegraphics[
natheight=2.537400in,
natwidth=3.044100in,
height=2.5374in,
width=3.0441in
]%
{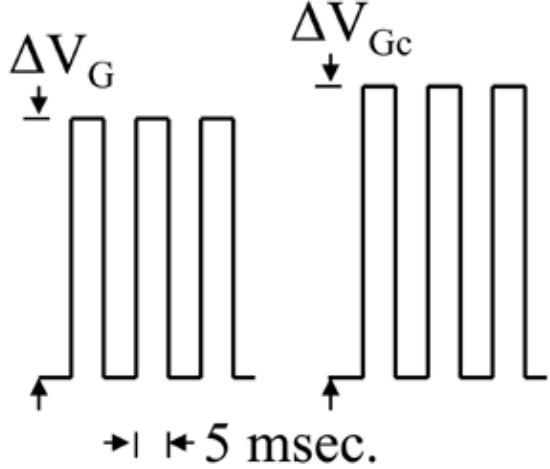}%
\caption{Square-wave pulse outputs from detector amplifier (not to scale) for
an initially depleted $\Phi_{G}$. The measured pulse height $\Delta V_{G}$
with $B_{R}$ blocking $\Phi_{R}$ increases to $\Delta V_{Gc}$ by equilibration
transfer of irradiance from $\Phi_{R}$ on the coupling path.}%
\label{f-cap-13}%
\end{center}
\end{figure}
After $\Delta V_{Gc}$ is acquired, $\Phi_{R}$ is blocked from the coupling
path by $B_{R}$. The detector then samples the same annular region of
$\Phi_{G}$ but now the peak height measurement%
\begin{equation}
\Delta V_{G}=\kappa E_{G}%
\end{equation}
provides the beam's grating-emergent energy $E_{G}$, unmodified by coupling,
to within the same multiplicative constant $\kappa$. The vital significance of
these two detector measurements is that their ratio%
\begin{align}
\frac{\Delta V_{G}}{\Delta V_{Gc}}  &  =\frac{E_{G}}{P_{G}}\nonumber\\
&  =\Omega_{G\text{ }ex}%
\end{align}
which is the experimentally determined occupation value from a single pair of
measurements $\Delta V_{Gc}$ and $\Delta V_{G}$. The subscript
\textquotedblleft$ex$\textquotedblright\ has been added to clearly identify
this quantity as experimentally determined. The final averaged $\Omega
_{G\text{ }ex}$ reported in the next subsection for each of the three gratings
studied are computed from multiple Eq. (51) determinations which are
themselves averages of $\sim100$ pulse cycles.

The theoretical counterpart to those final $\Omega_{G\text{ }ex}$ experimental
values is the Eq. (41) function%
\begin{align}
\Omega_{r;0.5}(\alpha_{t})  &  =\Omega_{r;0.5}(\pi w/\lambda)\nonumber\\
&  \equiv\Omega_{G[j(w)]\text{ }th},
\end{align}
shown in Fig. 7 and re-designated here as $\Omega_{G[j(w)]\text{ }th}$ since
the wavelength is invariant at $\lambda=633$ $nm$, $G$ is uniquely
characterized by the continuum value $j(w)$, and the added subscript
\textquotedblleft$th$\textquotedblright\ explicitly emphasizes the theoretical
basis of the function.

In the foregoing method for calculating $\Omega$ from experimental
measurements, it is necessary to cast the equations in terms of the directly
detectable beam energies (as derived from beam powers) in order to infer the
beam probabilities which are themselves not readily amenable to direct
measurement. However, it is probability creation and annihilation that is of
fundamental significance to duality violation in LRQM and therefore it is
important from a theoretical perspective to explicitly express the resultant
occupation value entirely in terms of probability.

For the general conditions of an ordinary output%
\[
\Omega_{o}=\frac{E_{o}}{P_{o}}=1
\]
and a resultant%
\[
\Omega_{r}=\frac{E_{r}}{P_{r}}.
\]
With energy conservation%
\[
E_{o}=E_{r}%
\]
in the transition and identifying%
\begin{equation}
P_{r}=P_{o}\pm\Delta P,
\end{equation}
we have the equation%
\begin{equation}
\Omega_{r}=\left(  1\pm\frac{\Delta P}{P_{o}}\right)  ^{-1}%
\end{equation}
where a positive sign represents $\Delta P$ probability creation and a
negative sign represents $\Delta P$ probability annihilation.

\subsection{Experimental results}

Consistent with the notation in Eq. (52), the experimentally determined
$\Omega_{G[j(w)]\text{ }ex}$ values specific to the three gratings are
$\Omega_{G(2.63)\text{ }ex}$, $\Omega_{G(3.16)\text{ }ex}$, and $\Omega
_{G(3.94)\text{ }ex}$.

The three experimental values $\Omega_{G\text{ }ex}$ are plotted for
comparison on the Fig. 7 theoretically predicted $\Omega_{G[j(w)]\text{ }th}$.
The experimental values $\Omega_{G(2.63)\text{ }ex}=1.015\pm0.003$ with a
duality modulation of $+1.5\%$, $\Omega_{G(3.16)\text{ }ex}=0.982\pm0.003$
with a duality modulation of $-1.8\%$, and $\Omega_{G(3.94)\text{ }%
ex}=0.998\pm0.003$ with a duality modulation of $-0.2\%$ are in good agreement
with the theoretically predicted $\Omega_{G[j(w)]\text{ }th}$ function for
LRQM shown in Fig. 7. At the most fundamental level, quite independent from
the LRQM theoretical basis presented here, any experimentally significant
$\Omega_{G[j(w)]\text{ }ex}$ deviations from unity demonstrate net transfers
of energy during coupling that are inconsistent with PIQM.

\subsection{Analysis of biased error}

We conclude this section with an analysis of the three idealized coupling
criteria identified in subsection D:

(1) The coupling path has perfect equilibration efficiency.

(2) $\Phi_{R}$ serves as an infinite source or sink.

(3) $\Phi_{R}$ is totally excluded from the detector sampling region by the
mask on $B_{m}$.

As a practical matter these criteria are not fully achieved in the Figs. 10
and 11 apparatus and, accompanying the usual statistical dispersion of
measured values, there are biased (non-random) sources of error present
related to these criteria. We show below that these sources cause the
experimentally measured $\Omega_{G\text{ }ex}$ to underestimate the duality
modulation, i.e. the actual magnitudes of the duality violations are larger
than that of the current experimentally measured duality modulations.
Moreover, we show that the magnitudes of the underestimates are not a
significant fraction of their respective duality modulations. Despite the
smallness of these underestimates, we include the biased error analysis here
as an exercise in completeness and in the interests of identifying how the
measurements can be optimized. These criteria and calculation of $\Omega$
underestimates are examined in greater detail in ref. \cite{mirell2}.

Criterion (1) relates to an idealized coupling of two beams that fully
equilibrates them to a common $\Omega$. This equilibration is mediated by the
physical proximity of the two beams and the longitudinal extent of that
proximity. Both of these factors are addressed in the current experimental
configuration by the coaxial propagation of these beams on a 2000 mm coupling
path and the iris diameter.

Matching of the respective beam Gaussian diameters along the coupling path is
necessarily altered by deliberate convergence of $\Phi_{R}$ onto the $B_{m}$
mask in order to provide substantial separation of that beam from the detector
annular sampling region. Nevertheless, this configuration still provides for
effective coupling over the 2000 mm path. Longer coupling paths have been
examined in the interests of potentially improving the coupling efficiency,
but with no significant change in $\Omega_{G\text{ }ex}$ observed, the 2000 mm
path was retained.

Similarly, for the Fig. 11 beam configuration, coupling path efficiency is
further optimized by maximizing radial equilibration between $\Phi_{G}$ and
$\Phi_{R}$. This is achieved by confining the sampled $\Phi_{G}$ to an annular
region most closely coupled to the convergent $\Phi_{R}$ using a minimal iris
setting (that still admits sufficient $\Phi_{G}$ beam power for measurement).

In any case, an incomplete equilibration arising from a non-ideal coupling
implies that the $\Delta E$ transfer is less than it would be if $\Omega
_{Gc}\rightarrow1$. As a result, the apparent $P_{G}$ equated to the measured
$E_{Gc}$ would be underestimated for depletion and overestimated for
enrichment yielding an underestimated magnitude of duality modulation for both.

In order to demonstrate the magnitude of the duality modulation underestimate
for the not fully achieved criteria (2) and (3), it is most useful to begin
with idealized experimental values $\Omega_{G(3-)\text{ }ex,id}=1.025$
and\ $\Omega_{G(3+)\text{ }ex,id}=0.975$ for which criteria (2) and (3) are
achieved in principle and calculate the respective apparent values
$\Omega_{G(3\pm)\text{ }ex,ap}$.

For criterion (2), if $\Phi_{R}$ has only a finite probability $P_{R}$,
perfect coupling path equilibration from criterion (1) does still provide
equalized final occupation values%
\begin{equation}
\Omega_{G(3\pm)c\text{ }ex}=\Omega_{Rc\text{ }ex,id}\neq1
\end{equation}
but because of the finite $P_{R}$, the initial $\Omega_{G(3\pm)\text{ }ex,id}$
and $\Omega_{R}$ are mutually convergent on a non-unit value. For the
intensive $\Omega$'s and extensive $P$'s and $E$'s, it can readily be shown
that%
\begin{align}
\Omega_{G(3\pm)c\text{ }ex}  &  =\frac{E_{G}+E_{R}}{P_{G}+P_{R}}\nonumber\\
&  =\Omega_{G\text{ }ex,id}\frac{P_{G}}{P_{G}+P_{R}}+\Omega_{R}\frac{P_{R}%
}{P_{G}+P_{R}}%
\end{align}
where, for the initially ordinary $\Phi_{R}$, $E_{R}=P_{R}$ and $\Omega_{R}%
=1$.\cite{mirell2}

To examine this expression in the context of the present experiment, where
$E_{R}\approx100E_{G}$, we note that the probabilities are also approximately
related by $P_{R}\approx100P_{G}$ despite a small inequality of $E_{G}$ and
$P_{G}$ for a small duality modulation of the initial $\Phi_{G}$. These
energies, which are measurable by detector, may be substituted for the
respective probabilities in the coefficients $P_{G}/(P_{G}+P_{R})$
and$\ P_{R}/(P_{G}+P_{R})$. This substitution introduces only a second order
error in Eq. (56) since the respective $P\approx E$. Then%
\begin{equation}
\Omega_{G(3\pm)c\text{ }ex}=\frac{1}{101}\Omega_{G(3\pm)\text{ }ex,id}%
+\frac{100}{101}.
\end{equation}
For $\Omega_{G(3-)\text{ }ex,id}=1.025$,%
\begin{align}
\Omega_{G(3-)c\text{ }ex}  &  =1.0002\nonumber\\
&  =\Omega_{G(3-)c\text{ }ex,id}\nonumber\\
&  =\frac{E_{Gc}}{P_{G}}\nonumber\\
&  =\Omega_{Rc\text{ }ex}.
\end{align}
Then, for the finite $\Phi_{R}$ that does not yield a fully realized criterion
(2), $E_{Gc}=1.0002P_{G}$ is treated as $P_{G}$ and the apparent
experimentally measured occupation value%
\begin{equation}
\Omega_{G(3-)\text{ }ex,ap}=\frac{E_{G}}{E_{Gc}}=\frac{E_{G}}{1.0002P_{G}%
}=\frac{1}{1.0002}\Omega_{G(3-)\text{ }ex,id}.
\end{equation}
From this result we find that the apparent $\Omega_{G(3-)\text{ }ex,ap}$ is
insignificantly smaller than the idealized $\Omega_{G(3-)\text{ }ex,id}$ and
the apparent duality modulation $\Omega_{G(3-)\text{ }ex,ap}-1$ is also
insignificantly smaller than the idealized value.

Similarly, for $\Omega_{G(3+)\text{ }ex,id}$ the apparent $\Omega_{G(3+)\text{
}ex,ap}$ is insignificantly larger than the idealized $\Omega_{G(3+)\text{
}ex,id}$ and the magnitude of the apparent duality modulation $\left\vert
\Omega_{G(3+)\text{ }ex,ap}-1\right\vert $\ is also insignificantly smaller
than the idealized value.

Lastly, we consider deviation from criterion (3). In practice, the Gaussian
tail of the convergent $\Phi_{R}$ beam spot always has some small but finite
fraction of $P_{R}$ outside the $B_{m}$ mask into the annular sampling region.
As a consequence, $\Phi_{R}$ is not completely separated from the annularly
sampled $\Phi_{G}$. Conversely, that sampling accepts a substantial fraction
of $P_{G}$. The general significance of this incomplete separation relates
again to an apparent discrepancy in $\Delta E$ transferred to or from
$\Phi_{G}$ to produce an $E_{Gc}$ that is presumed to have equivalence to
$P_{G}$. In the equilibration process an equal $\Delta E$ is transferred
respectively from or to $\Phi_{R}$. Upon coupling, some large fraction $F_{G}$
of $\Delta E$ transfers to or from $\Phi_{G}$ in the annular region while
concurrently some small but finite fraction $F_{R}$ of $\Delta E$ transfers
respectively from or to $\Phi_{R}$ in the same annular region resulting in a
diminished net energy transfer in the annular measurement region.

This dependency of the occupation value on the portion of $\Phi_{R}$ in the
annular sampling region is given by an apparent%
\begin{equation}
\Omega_{G(3\pm)\text{ }ex,ap}=\frac{F_{G}E_{G}}{F_{G}E_{G}\pm(F_{G}\Delta
E-F_{R}\Delta E)}%
\end{equation}
where $\pm$ applies respectively to an initially depleted or enriched
$\Phi_{G}$.\cite{mirell2} To put this error into perspective with regard to
the apparatus of the present experiment, $F_{G}\approx0.4$ and $F_{R}%
\approx0.01$. For the idealized $\Omega_{G(3\pm)\text{ }ex,id}$, the magnitude
of the duality modulation is $\left\vert \Omega_{G(3\pm)\text{ }%
ex,id}-1\right\vert =\Delta E/E_{G}=0.025$. With these values, an idealized
$2.5\%$ magnitude of duality modulation for both depleted and enriched
$\Phi_{G}$ is again underestimated as an apparent insignificantly smaller magnitude.

We conclude that the close approximations to the criteria (1), (2), and (3)
for the beam parameters in the present experiment provide apparent
$\Omega_{G\text{ }ex,ap}$ values that differ insignificantly from idealized
$\Omega_{G\text{ }ex,id}$ values. Consequently, no compensating adjustment of
experimental results $\Omega_{G\text{ }ex}$ in the previous subsection is
required with regard to the coupling criteria.

\section{Discussion}

A viable locally real alternative to the probabilistic interpretation of
quantum mechanics PIQM must necessarily be in agreement with performed
experiments and must provide a self-consistent theoretical basis for
representing quantum mechanical phenomena.

We hypothesize that a comprehensive locally real representation of an
underlying quantum mechanical formulation, LRQM, can be systematically
constructed by providing for separable locally real wave-like and
particle-like physical entities where the wave-like entity is represented on a
covariant field of elementary oscillators generally in ground state with
relative non-coherence.\cite{mirell} A photon in this representation consists
of a wave packet of these ground state oscillators in relative coherence, i.e.
the wave-like entity. The particle-like entity, i.e. the energy quantum,
consists of a raised energy state of one of these oscillators in coherent
motion. The density of these oscillators in relative coherence at a particular
location on the wave packet $\Phi$ is given by $\left\vert \Phi\right\vert
^{2}$. This density of oscillators in relative coherent motion represents a
small fraction of the total density of field oscillators at any given point.
Consistent with Born's rule, $\left\vert \Phi\right\vert ^{2}$ provides the
probability flux density for the location of the energy quantum on the wave
packet. For ordinary ($\Omega=1$) photons, the integration of $\left\vert
\Phi\right\vert ^{2}$ over all space (essentially, over the entire wave
packet) yields a value that is in strict proportion to the unit energy quantum
consistent with the PIQM principle of duality.

We emphasize that these LRQM interpretations, restricted to a simple system
such as an ordinary discrete photon, are not measurably distinguishable from
those of PIQM. Differentiability arises when the structure of the wave packet
is altered in such a way that local probability on a photon is not conserved.
For correlated entities, this is shown in ref. \cite{mirell} to be a
constraint on the ensemble of quantum states in Hilbert space that results in
a probability loss on one of the correlated entities. The locally real
solution, which is independent of Bell's Theorem \cite{bell}, is consistent
with the underlying quantum mechanical formalism and agrees with performed experiments.

In the present investigation of duality, non-conservation of probability is
again the factor distinguishing LRQM from PIQM. With respect to duality, the
probability non-conservation is manifested as an alteration of the local flux
density $\left\vert \Phi\right\vert ^{2}$ of oscillators in relative
coherence. At the simplest level, the wave packet of an ordinary discrete
photon traversing a beam splitter emerges as a pair of spatially similar wave
packets along the transmissive and reflective output channels but with
relative oscillator coherence densities reduced to $T\left\vert \Phi
\right\vert ^{2}$ and $R\left\vert \Phi\right\vert ^{2}$, respectively where
$T$ is the transmission factor and $R$ is the reflection factor of the beam
splitter. The energy quantum transfers onto one of these two packets, again
consistent with Born's rule. Probability is conserved when summed over both
output channels ($T+R=1)$ and PIQM retains duality by invoking a non-local
probabilistic (non-real) interpretation of this phenomenon. From the
perspective of LRQM, the two emergent packets, only one of which is occupied
by the energy quantum, both exhibit physical distinctions from the incident
photon. However, the testability of these distinctions at the discrete level
is problematic. This difficulty is circumvented here by going to the
continuous wave (cw) regime and using a grating system presumptively
generating duality-modulated beams since even very modest duality modulations
are definitively testable in the cw regime.

The LRQM treatment of probability as a relative quantity can be demonstrated
from the perspective of the densities of the field oscillators in coherent
motion on the resultant diffraction beams emergent from a grating near
threshold. For example, $\sim2.5\%$ of the output probability that forms the
resultant probability is annihilated in the near zone of grating $G(3-)$.
Equivalently, this translates to a $\sim2.5\%$ reduction in the coherent
oscillator density on each of the $G(3-)$ resultant orders. However, as the
output energy quanta in the near zone of $G(3-)$ transfer onto the resultants,
the proportionate $\sim2.5\%$ reduction in the coherent oscillator density on
each does not alter the relative distribution of the total quanta onto those
orders. Indeed, any proportionate reduction or increase in coherent oscillator
density on a complete set of resultant channels does not change the
distribution of a given set of quanta transferring onto those channels.
Probability, treated as a measure of the physical coherent oscillator density,
is appreciated as a relative quantity expressing the expectation of quanta
transferring to a particular channel.\cite{saxon}

Conversely, the motivation to compactly fold conservation of the particle-like
energy quantum in with the distributional probability results in a PIQM
profoundly distinctive from LRQM. The PIQM normalization of the integrated
$\left\vert \Phi\right\vert ^{2}$ over all space effectively elevates the
wave-like probability to equivalence with the particle-like energy quantum.
This is really the fundamental statement of duality. An objective review of
the distinctions imposed by PIQM duality is given by
Rabinowitz.\cite{rabinowitz}

In PIQM energy quanta and probability are then dual manifestations of a single
entity, the \textquotedblleft photon\textquotedblright. However, when
superposition states such as those created by a beam splitter are considered,
PIQM is forced to impose the properties of non-locality and non-reality which
particularly distinguish that interpretation from LRQM.

Alternatively, in the comprehensive LRQM that emerges, the field coherence
states are represented by the wave functions that exhibit probability
non-conservation for particular quantum phenomena. These wave functions, freed
from the constraint of duality-imposed renormalization, \textquotedblleft
complete\textquotedblright\ the underlying quantum mechanical formalism in the
regard that the quantum phenomena are then representable as locally real. In
this context, Einstein, Podolsky, and Rosen had referred to objective
properties of a physical system that are represented by parameters which they
called \textquotedblleft elements of reality\textquotedblright.\cite{einstein}
In common usage these elements have since, unfortunately, been re-identified
as \textquotedblleft hidden variables\textquotedblright. Ferrero, Marshall,
and Santos present a compelling argument for the inappropriateness of this
re-identification. Their argument has application here since the LRQM wave
function is shown to express objective properties before any measurement is
made on it as they prescribe in ref. \cite{ferrero}. Moreover, the objective
properties of that wave function can realized by physical measurement. The
intensity $\left\vert \Phi\right\vert ^{2}$ of a beam is disproportionate to
that beam's irradiance for a non-ordinary beam. That disproportion is
measurable from the energy quanta transfer that occurs during equilibration
coupling of that beam with an ordinary beam.

From the perspective of PIQM, for which probability is systematically
conserved in the preparation of the normalized wave function, the variables
that would provide locality appear to be hidden since phenomena that
distinguish PIQM from LRQM are associated with non-conservation of probability.

Ferrero, Marshall, and Santos postulate that \textquotedblleft in spite of the
spectacular success of quantum mechanics, it is worthwhile exploring (small)
modifications of the formalism in order to ensure compatibility with local
realism.\textquotedblright\ \cite{ferrero} This comment was made in the
support of seeking a locally real theory that naturally obviates entanglement,
but necessarily implies extension to an encompassing theory that also
naturally excludes quantum phenomena that invoke non-locality such as duality.
The objective posed by Ferrero, Marshall, and Santos is intrinsic to LRQM by
incorporating the underlying quantum formalism modified only by the omission
of probability renormalization in transitional processes. The resultant LRQM
is broadly consistent with PIQM predictions and performed experiments while
providing for a local reality-based representation.

\section{Conclusions}

The experiment reported here provides a highly reproducible violation of
quantum mechanical duality using an apparatus configured with two independent
HeNe lasers and readily available components. The determination of that
duality violation resolves to easily measured variations in continuous wave
laser beam power, variations predicted by a locally real representation of
quantum mechanics but excluded by the probabilistic interpretation of quantum mechanics.

\end{document}